\documentclass[prc,twocolumn,showpacs]{revtex4}
\usepackage[dvips]{graphicx}
\usepackage{color}
\usepackage{latexsym}
\usepackage{amsmath}
\usepackage{amssymb}
\newcommand{\nph}[1]{#1$p$#1$h$}

\newcommand{\ket}[1]{| #1 \rangle}
\newcommand{\bra}[1]{\langle #1 |}

\newcommand{\dmom}[1]{\int \frac{\mathrm{d}^3 #1}{(2\pi)^3} \ }
\newcommand{\dpos}[1]{\int \mathrm{d}^3 #1 \ }

\newcommand{\dotp}[2]{\langle #1 | #2 \rangle}
\newcommand{\meh}[2]{\langle #1 | {\mathbf H} | #2 \rangle}

\newcommand{\me}[3]{\langle #1 | #2 | #3 \rangle}

\newcommand{\msa}[2]{S_{\vec{#1} #2} }
\newcommand{\msb}[2]{S_{#1 #2} }

\newcommand{\mta}[2]{\tilde S_{\vec{#1} #2} }
\newcommand{\mtb}[2]{\tilde S_{#1 #2} }

\newcommand{\Op}{{\mathcal O}}

\newcommand{\h}{{\mathbf H}}
\newcommand{\ke}{{\mathbf T}}
\newcommand{\vo}{{\mathbf V_0}}

\newcommand{\s}{{\mathbf S}}
\newcommand{\pcm}{{\mathbf P}}

\newcommand{\no}[2]{\left ( #1 \right )_{#2}}

\newcommand{\Ond}[1]{{\mathbf O}^\dag_{#1}}
\newcommand{\On}[1]{{\mathbf O}_{#1}}

\newcommand{\Cd}[1]{{\mathbf c}^\dag_{#1}}
\newcommand{\C}[1]{{\mathbf c}_{#1}}
\newcommand{\Ad}[1]{{\mathbf a}^\dag_{\vec #1}}
\newcommand{\A}[1]{{\mathbf a}_{\vec #1}}
\newcommand{\Bd}[1]{{\mathbf b}^\dag_{#1}}
\newcommand{\B}[1]{{\mathbf b}_{#1}}
\newcommand{\CAdBd}[3]{( + \, {\mathbf b}_{#3}^\dag \, \delta_{\vec{p_#1} \vec{#2}} ) }
\newcommand{\CdAdBd}[3]{( - \, {\mathbf a}_{\vec #2}^\dag \, \delta_{n_#1 #3} ) }

\begin{document}
%
%
%
\preprint{LA-UR-03-4197}
\title[continuum]
   {Continuum coupled cluster expansion}
\author{Bogdan Mihaila}
\affiliation{Theoretical Division,
   Los Alamos National Laboratory,
   Los Alamos, NM 87545}
\email{bmihaila@lanl.gov}
\date{\today}
\begin{abstract}
We review the basics of the coupled-cluster expansion formalism
for numerical solutions of the many-body problem, and we outline
the principles of an approach directed towards an adequate
inclusion of continuum effects in the associated single-energy
spectrum. We illustrate our findings by considering the simple
case of a single-particle quantum mechanics problem.
\end{abstract}
\maketitle
%
%

\section{Introduction}
\label{sec:intro}

One of the major endeavors in modern nuclear physics, is the
ongoing quest for an effective nuclear many-body theory, a key
element in the attempt to extrapolate existing experimental data
to regimes which are not currently accessible in the laboratory.
Ideally, one would like to begin with the experimental data
available from nucleon-nucleon ($N\!N$) scattering
experiments~\cite{ref:stoks_1990,ref:stoks_1993a} and determine
the underlying $N\!N$
interaction~\cite{ref:stoks_1994,ref:stoks_1993b,ref:av18,ref:rupert_1996,ref:rupert_2001,ref:rupert_2002}.
Subsequent nuclear structure calculations are then
necessary to determine a three-nucleon 
interaction~\cite{ref:il2}, to account for the remaining
differences between theory and experiment.

Direct comparison of theory (calculations) to experimental data is
ambiguous if the nuclear many-body problem is not solved
accurately. Results of numerical calculations can be used to
constrain fundamental aspects of the theory, provided that one has
good approximations of the exact result in order to avoid
introducing a bias due to the approximations involved. Much
progress has been achieved recently by the Green's Function Monte
Carlo (GFMC)~\cite{ref:carlson_1987} collaboration in obtaining
accurate description of nuclear structure properties of light
nuclei ($A \le 10$)~\cite{ref:PW01,ref:PVW02}. Such ab initio
calculations, based on realistic interaction models are key for
understanding in a quantitative manner the experimental data
provided by modern accelerator facilities at Jefferson Laboratory
(JLAB) and the future Rare Isotope Accelerator (RIA). We believe
that a crucial undertaking in modern physics involves the
extension of this work to heavy nuclei.

The limitations of the GFMC method to carry out calculations for
arbitrary size nuclei, originate both from the inherent Fermi sign
problem associated with a non-relativistic system of nucleons
obeying Fermi statistics, and the fact that in the present
implementation of the GFMC one carries out explicit summations
over the spin-isospin degrees of freedom. As such, the relevant
number of degrees of freedom increases as $2^A (^A_Z)$, and the
size of the spin-isospin vector space quickly becomes prohibitive.
A possible way around this technical problem, is offered by the
Auxiliary Field Diffusion Monte Carlo (AFDMC)
method~\cite{ref:kevin_1999}, where one attempts to sample the
spin-isospin degrees of freedom in addition to the spatial
coordinates of the nucleons. This method is currently under
development, and encouraging results have been recently reported
for neutron matter~\cite{ref:kevin_2001}.

For nuclei  with A$\le$4, the nonrelativistic Schr\"odinger
equation with realistic nuclear interactions can be solved very
precisely: In a recent benchmark~\cite{ref:kamada02} carried out
for a $^4$He-like nucleus using the Argonne v$_8$' interaction
(central, spin-spin, tensor, spin-orbit and the corresponding
spin-exchange counterparts), the binding energy calculated using
seven ``exact'' many-body formalisms agreed within 0.5\%.
Unfortunately, few many-body formalisms are currently capable to
offer accurate numerical solutions of the nuclear many-body
problem with realistic interactions, when heavier systems are
concerned. For finite nuclei with $A>4$, the no core shell model
(NCSM)~\cite{ref:peter_2000a,ref:peter_2000b,ref:peter_2002} and
the coupled-cluster expansion (CCE) are the only methods which are
able to directly compare results with the GFMC for the same type
of hamiltonian. Both methods however have serious limitations
related to a slow convergence of the results with the size of the
model space. As we are about to describe in this paper, the
shortcomings of the CCE are not inherent in nature, and we like to
believe that the approach to solving the CCE we propose here, will
allow the CCE to provide accurate theoretical output for a medium
to large domain of masses, in the framework of realistic nuclear
interaction~\cite{ref:av18,ref:uix,ref:il2} and nuclear
currents~\cite{ref:CS98}. We call this new approach the
\emph{continuum} CCE (c-CCE), because continuum effects are
accurately taken in account.

The coupled-cluster expansion (CCE), also called the $\exp({\bf
S})$ method, was developed in the early 1960s by Coester and
K\"ummel~\cite{ref:C58,ref:CK60}. While the method is viewed as
exact, approximations are introduced stemming from truncations in
the CCE equations, as well as truncations in the model space.
Practical approaches for nuclear structure applications have been
notoriously difficult to realize. It was not until the 1970s that
Zabolitsky and K\"ummel~\cite{ref:Kummel_etal} were able to carry
out the first detailed calculations for finite nuclei, using a
representation of the wave function in coordinate space together
with common interactions of the time. The results for the binding
energy of nuclei such as $^4$He, $^{16}$O and $^{40}$Ca, were well
above the Coester line and were taken as evidence for the presence
of $3N$ and higher-order interactions.

While the CCE was extensively used in other areas in physics and
chemistry (see Refs.~\cite{ref:bishop_review,ref:navarro_review}
for additional information), further applications to nuclear
structure calculations based on realistic nuclear interactions
proved difficult to achieve. In the 1990s we mainly note the
attempt of constructing a translationally invariant CCE in
coordinate
space~\cite{ref:bishop_1990,ref:bishop_1996,ref:bishop_1998},
method which might yet offer an attractive alternative for finite
nuclei calculations (see
Refs.~\cite{ref:bishop_2000,ref:bishop_2002} for information on
recent progress on this approach.) To date, however, these
attempts are confined to semi-realistic interactions with a V4
structure (scalar plus spin-, isospin-, and space-exchange terms),
and the wave functions are not sophisticated enough to provide the
realistic account of nuclear forces required in the context of
realistic nuclear interactions, such as the Argonne v$_{18}$ and
corresponding three-nucleon forces.

Motivated by the availability of more sophisticated $N\!N$
interactions, and riding the wave of the ongoing expansion in
computer power, Heisenberg and
Mihaila~\cite{ref:bm_I,ref:bm_PRL,ref:bm_II} reexamined the
coupled-cluster approach and applied it to the spherical nucleus
$^{16}$O, together with a realistic nuclear interaction and
currents which were purposely consistent with the work done by the
GFMC collaboration: Calculations based on a realistic two-body
hamiltonian (Argonne v$_{14}$ and v$_{18}$) 
were reported in Ref.~\cite{ref:bm_I}, while effects of a
phenomenological three-body interaction (Urbana IX) were reported
in Ref.~\cite{ref:bm_II}. An extension of the later application,
involved the microscopic calculation of the charge form factor in
$^{16}$O. Results were presented in Ref.~\cite{ref:bm_PRL}, and
showed good agreement with the available experimental
data~\cite{ref:sick_o16} when both three-body forces and
meson-exchange currents were taken into account.

In its present implementation the CCE equations are solved for the
special case of spin-isospin shell-saturated nuclei, such as
$^{4}$He, $^{12,14}$C, $^{14,16}$O. In order to depart from the
closed-shell case one resorts to an equation of motion approach,
using the ground-state of the closed-shell nucleus as a new
vacuum. As such one can describe the properties of hole-state
nuclei such as $^{13}$C and $^{15}$N, and preliminary results are
promising. However, such an approach will inevitably compound the
effect of new and old approximations. Also, the present approach
has limitations due to the intrinsic cutoffs one has to impose
when defining the model configuration space. The single particle
states are expanded in a harmonic oscillator basis, and satisfy
the same type of boundary conditions for both hole and particle
states. In effect we discretize the continuum part of the
self-consistent one-body mean-field hamiltonian used to define the
single-energy spectrum. This results in the necessity of a large
$50 \hbar \Omega$ configuration space and subsequent significant
storage problems and lengthy execution time.



One would like to obtain exact solutions for the 2-, 3- and 4-body
systems via CCE. Since contributions due to $n>A$ particle-hole
configurations are always cancelled exactly, the above few-body
systems will provide scenarios when the truncation of the CCE
hierarchy for $A= 2$, $3$ and $4$, respectively, becomes exact,
and we will be able to reliably ascertain the numerical accuracy
by direct comparison with similar results obtained via other exact
methods. Moreover, by solving the two- and three-body problem, we
will effectively eliminate our present reliance on the
closed-shell hypothesis, and allow for direct calculations of
nuclei inside the shell via CCE. Given our past experience
regarding the ground state calculation of spin-isospin
shell-saturated nuclei in configuration space, we can expect that
a four-cluster truncation of the CCE equations will allow an
accurate description of arbitrary nuclei beyond $A=16$.


From a numerical standpoint, we must always remember that we have
to solve a finite physical system, using finite computational
resources, in a finite amount of time. We submit that the
resolution of the many-body problem is linked to the successful
design of a numerical algorithm that allows for an efficient
implementation on a massively multiprocessor machines. In order to
preserve the scalability of a numerical algorithm on a
multiprocessor machine, we propose to address the problem of
finding the energy spectrum of the hamiltonian by strictly
following the two commandments:
\begin{enumerate}
   \item Thou shalt \textbf{not} solve matrix eigenvalue equations.
   \item Thou shalt \textbf{not} perform explicit matrix inversions.
\end{enumerate}
In order to solve this conundrum, we must pursue those iterative
numerical schemes which are most likely approachable using Monte
Carlo techniques. It is our contention that the basic expertise
for solving the nuclear many-body problem is available at this
time: The GFMC is the only many-body method available today that
obeys the above constraints. However, in its present formulation,
the GFMC carries out explicit summations over all possible
spin-isospin combinations, and this leads to a $2^A (^A_Z)$
increase for the number of the relevant degrees of freedom. We
propose to investigate the possibility of solving a comparatively
small ($n\le 4$)-cluster truncation of the CCE hierarchy of
equations using Monte Carlo techniques. Assuming the CCE truly
exhibits a $1/A$-type convergence, then it is reasonable to expect
that approximations based on the CCE will accurately describe
nuclear structure with realistic nuclear interactions and
currents, around and beyond $^{16}$O. The CCE equations will still
involve explicit spin-isospin summations, but these will entail
only minimal complications because the small size of the cluster
will drastically simplify the problem.

To summarize, our long-range interest is two-fold: firstly, we
want to develop a continuum version of the CCE in an attempt to
efficiently describe physics at large distances; secondly, we are
interested in finding an efficient numerical procedure to solve
the CCE equations. We find it useful to begin our investigations
with the simplest dimensional realization of the many-body
problem: the one-body case, $A=1$. The mere simplicity of the
problem will allow for an exhaustive, detailed study of the CCE
equations.

\section{Single-energy spectrum}
\label{sec:spectrum}

Consider the Hilbert space of states associated with the one-body
hamiltonian
\begin{equation}
   \h_0 = \ke + \vo
   \>,
\end{equation}
and a basis set in this Hilbert space. In general, this basis
features both a discrete and a continuum part, such that any state
in the Hilbert space can be expanded as
\begin{equation}
   \ket{\psi} = \sum_n a_n \ket{\phi_n} + \dmom{p} b_{\vec p}
   \ket{\phi_{\vec p}} \>.
\end{equation}
That is to say that the basis states in the spectrum satisfy the
orthonormality relations
\begin{subequations}
\label{eq:orthonorm}
\begin{align}
   \dotp{\phi_n}{\phi_m} & = \delta_{nm}
   \>,
\label{eq:discrete}
   \\
   \dotp{\phi_{\vec p}}{\phi_{\vec{p '}}} & = (2\pi)^3 \, \delta(\vec p - \vec {p '})
   = \delta_{\vec p \vec{p '}}
   \>,
\label{eq:cont}
   \\
   \dotp{\phi_n}{\phi_{\vec p}} & = 0
   \>,
\label{eq:mix}
\end{align}
\end{subequations}
and the basis set is complete
\begin{equation}
   \sum_n \ket{\phi_n}\bra{\phi_n} + \dmom{p}
   \ket{\phi_{\vec p}}\bra{\phi_{\vec p}} = \delta(\vec r - \vec {r '})
   \>.
\label{eq:complete}
\end{equation}
The above basis can be the same as, but is not restricted to, the
set of eigenvectors corresponding to the single-energy spectrum
defined by the solutions of the Schr\"odinger equation
\begin{equation}
   \h_0 \ \ket{\phi} \ = \ \varepsilon \ \ket{\phi}
   \>.
\end{equation}

We propose to start with a \emph{known} set of discrete single
particle wave functions $\{ \phi_n \}$ in the above Hilbert space,
which satisfy Eq.~(\ref{eq:discrete}), and subsequently construct
a set $\{ \phi_{\vec p}\}$, which obeys the remaining
orthonormality conditions, Eqs.~(\ref{eq:cont},\ref{eq:mix}), the
closure relation (\ref{eq:complete}), and satisfies a
predetermined boundary condition (i.e. bound state or scattering
state).

This can be done as follows: In order to satisfy the orthogonality
conditions between the appropriate continuum single-particle wave
functions, we start with a complete set of continuum wave
functions
\begin{equation}
   \ket{\bar \chi_{\vec p}} = \ket{\chi_{\vec p}}
   - \sum_n  \ket{\phi_n} \dotp{\phi_n}{\chi_{\vec p}}
   \>.
\label{eq:overlap}
\end{equation}
For instance, if the continuum wave function $\chi_{\vec p}(\vec
r)$ is just a plane wave, $\chi_{\vec p}(\vec r)=e^{\mathrm{i}\,
\vec p \cdot \vec r}$, then $\dotp{\phi_n}{\chi_{\vec p}}=\tilde
\phi_n(\vec p)$ denote the Fourier transform of the wave function
$\phi_n(\vec r)$. The set of continuum wave functions $\{ \bar
\chi_{\vec p} \} $ are orthogonal to the discrete spectrum of the
hamiltonian, i.e. $\dotp{\phi_n}{\bar \chi_{\vec p}} = 0$, and
satisfy the completeness relation
\begin{equation}
   \sum_n \ket{\phi_n}\bra{\phi_n} + \dmom{p}
   \ket{\bar \chi_{\vec p}}\bra{\bar \chi_{\vec p}} = \delta(\vec r - \vec {r '})
   \>.
\end{equation}
These wave functions however are not orthogonal to each other
\begin{equation}
   \dotp{\bar \chi_{\vec p}}{\bar \chi_{\vec{p '}}}
   = \delta_{\vec p \vec{p '}}
   - \sum_n \tilde \phi_n^*(\vec p) \ \tilde \phi_n(\vec {p '})
   \>.
\end{equation}
Without loss of generality we can orthonormalize this set $\{ \bar
\chi_{\vec p} \}$ to obtain the desired set of continuum wave
functions, $\{ \phi_{\vec p} \}$. Note that there is no
inconsistency here, since $\bar \chi_{\vec p}(\vec r)$ is not
simply a linear superposition of plane waves, but we can define
$\phi_{\vec p}(\vec r)$ as a linear combination of $\bar
\chi_{\vec p}(\vec r)$ in order to perform the orthonormalization.
In practice, the fact that the definition of $\bar \chi_{\vec
p}(\vec r)$ involves the Fourier transform of $\phi_n(\vec r)$,
establishes the appropriate grid representation of the
single-particle wave functions. The set $\{ \phi_{\vec p} \}$ is
obtained by orthonormalization of the set $\{ \bar \chi_{\vec p}
\}$ at these grid points. This concludes our construction.

The fact we actually have to normalize the continuum basis states
on a grid, may be a serious drawback. After all we started this
process with the idea in mind that a formal discretization of the
continuum prevents us from obtaining rigorous numerical
convergence with modest computational resources. It is therefore
unfortunate that an actual discretization appears to be necessary
at this point. However, we claim that already at this point we are
better off in this formalism since \emph{all} basis states, and
most importantly the continuum basis states, have the correct
boundary conditions built in. Moreover, we will see that the above
construction 
is not binding us into a certain course of action. We will argue
later, that we can in fact circumvent these apparent difficulties,
and return to a representation formulated entirely in coordinate
space where the explicit knowledge of the continuum states is not
mandatory.

\section{Second quantization}
\label{sec:secq}

We proceed now by outlining the second quantization language we
will subsequently use. We start by introducing the field operators
$\mathbf{ \hat \psi}(\vec r)$ and $\mathbf{ \hat \psi}^\dag(\vec
r)$, as linear combinations of creation and destruction operators
$\C{\alpha}, \Cd{\alpha}$, respectively:
\begin{align}
   \mathbf{ \hat \psi}(\vec r)
   & = \sum_\alpha \phi_\alpha (\vec r) \ \C{\alpha}
   \\ \notag
   & \equiv
   \sum_n \ \phi_n (\vec r) \ \Bd{n}
   \ + \
   \dmom{p} \phi_{\vec p} (\vec r) \ \A{p}
   \>,
\end{align}
where the single-particle wave functions $\phi(\vec r) \equiv \{
\phi_n (\vec r), \phi_{\vec p} (\vec r) \}$, form a complete
basis, and satisfy the orthonormality properties, as explained in
the previous section. The inverse transformations are
\begin{align}
   \Bd{n} & =
   \dpos{r} \phi_n^* (\vec r) \, \mathbf{ \hat \psi}(\vec r)
   \>,
   \\
   \A{p} & =
   \dpos{r} \phi_{\vec p}^* (\vec r) \, \mathbf{ \hat \psi}(\vec r)
   \>.
\end{align}
In other words, $\Bd{n}$ removes a nucleon from the \emph{hole}
orbit $\phi_n(\vec r)$, while $\A{p}$ removes a nucleon from a
\emph{particle} orbit $\phi_{\vec p}(\vec r)$. With these
definitions, we can write
\begin{gather}
    {\mathbf c}^\dag =    \left \{
       \begin{matrix}
          \Ad{p} \>, & \mathrm{if}\ \varepsilon_p \ge \varepsilon_F \>, \\
          \B{n} \>, & \mathrm{if}\ \varepsilon_n < \varepsilon_F \>,
       \end{matrix}
    \right .
\end{gather}
and
\begin{gather}
    {\mathbf c} =    \left \{
       \begin{matrix}
          \A{p} \>, & \mathrm{if}\ \varepsilon_p \ge \varepsilon_F \>, \\
          \Bd{n} \>, & \mathrm{if}\ \varepsilon_n < \varepsilon_F \>.
       \end{matrix}
    \right .
\end{gather}
Here, $\varepsilon_F$ denotes the Fermi energy. Note that in the
region between the Fermi energy and the separation energy
$(\varepsilon_p = 0)$, the ``continuum'' spectrum,
$\ket{\phi_{\vec p}}$, is actually discrete, and the integral has
to be interpreted as a sum.

Since we are interested in many-body systems obeying fermi
statistics, we ask the field operators to obey the
anti-commutation relations
\begin{equation}
   \Bigl \{ \mathbf{ \hat \psi}^\dag (\vec r), \,
           \mathbf{ \hat \psi} (\vec {r '}) \Bigr \} =
   \delta(\vec r - \vec {r '}) \>.
\end{equation}
Spin and isospin indices are implied. In turn, the creation and
annihilation operators obey the canonical relations:
\begin{subequations}
\label{eq:Canti}
\begin{gather*}
    \{ \Cd{n}, \C{m} \} = \delta_{nm} \>, \qquad
    \{ \Cd{\vec p }, \C{\vec{p '}} \} = \delta_{\vec p \vec{p '}} \>,
    \\ \notag
    \{ \C{n}, \C{m} \} =
    \{ \Cd{n}, \Cd{m} \} =
    \{ \C{\vec p }, \C{\vec{p '}} \} =
    \{ \Cd{\vec p }, \Cd{\vec{p '}} \} = 0 \>,
    \\ \notag
    \{ \C{n}, \C{\vec p } \} =
    \{ \Cd{n}, \Cd{\vec p } \} =
    \{ \Cd{n}, \C{\vec p } \} =
    \{ \C{n}, \Cd{\vec p} \} = 0
    \>,
\end{gather*}
\end{subequations}
or
\begin{subequations}
\label{eq:ABanti}
\begin{gather*}
    \{ \Bd{n}, \B{m} \} = \delta_{nm} \>, \qquad
    \{ \Ad{p}, \A{p '} \} = \delta_{\vec p \vec{p '}} \>,
    \\ \notag
    \{ \B{n}, \B{m} \} =
    \{ \Bd{n}, \Bd{m} \} =
    \{ \A{p}, \A{p '} \} =
    \{ \Ad{p}, \Ad{p '} \} = 0 \>,
    \\ \notag
    \{ \B{n}, \A{p} \} =
    \{ \Bd{n}, \Ad{p} \} =
    \{ \Bd{n}, \A{p} \} =
    \{ \B{n}, \Ad{p} \} = 0 \>.
\end{gather*}
\end{subequations}

More definitions are in order: We introduce the \emph{bare}
vacuum, $\ket{0}$, such that
\begin{equation}
   \mathbf{\hat \psi}(\vec r) \, \ket{0} =0
   \>,
\end{equation}
or
\begin{equation}
   \A{p} \, \ket{0} =0
   \>,
   \qquad
   \Bd{n} \, \ket{0} =0
   \>.
\end{equation}
In turn, the \emph{physical} vacuum, $\ket{\Phi}$, which will play
the role of the reference state for $\exp(\s)$, is defined as the
exhaustive collection of minimal configurations of a given
symmetry, obtainable from the bare vacuum:
\begin{equation}
   \A{p} \, \ket{\Phi} = 0
   \>,
   \qquad
   \B{n} \, \ket{\Phi} = 0
   \>.
\end{equation}
We have
\begin{equation}
   \Cd{\vec p } \ket{\Phi} =
   \Ad{p} \ket{\Phi}
   \>,
   \qquad
   \C{n} \ket{\Phi} =
   \Bd{n} \ket{\Phi}
   \>.
\end{equation}

Finally, the second quantization representation of the one-body
hamiltonian operator is given by
\begin{equation}
   \h \ = \ \sum_{\alpha \beta}
   \Cd{\alpha} \, \meh{\phi_\alpha}{\phi_\beta} \, \C{\beta}
   \>.
\end{equation}
By the same token, we will represent for instance, the \nph{1}
cluster-correlation operator, as
\begin{equation}
   \s_1 \ = \ \sum_{n} \ \dmom{p}
              \Ad{p} \, \msa{p}{n} \, \Bd{n} \>.
\end{equation}

\section{CCE equations}
\label{sec:CCE}

In a traditional shell-model approach, one calculates the matrix
elements of the hamiltonian
\begin{eqnarray}
   \meh{\Psi_1}{\Psi_2} & = &
   \sum_{mn} a_n^{(1)*} a_m^{(2)} \meh{\phi_n}{\phi_m}
   \\ \nonumber &&
   + \sum_m \dmom{p} b_{\vec p}^* a_m^{(2)} \meh{\phi_{\vec p}}{\phi_m}
   \\ \nonumber &&
   + \sum_n \dmom{p '} a_n^{(1)*} b_{\vec{p '}} \meh{\phi_n}{\phi_{\vec{p '}}}
   \\ \nonumber &&
   + \dmom{p} \dmom{p '} b_{\vec p}^* b_{\vec{p '}} \meh{\phi_{\vec p}}{\phi_{\vec{p '}}}
   \>.
   \label{eq:heisy_1}
\end{eqnarray}
Then, for an arbitrary eigenstate of the hamiltonian,
$\ket{\Psi}$, we write the Schr\"odinger equation
\begin{equation}
   \h \ \sum_\alpha \ket{\Psi_\alpha}\dotp{\Psi_\alpha}{\Psi} =
   E \ \sum_\alpha \ket{\Psi_\alpha}\dotp{\Psi_\alpha}{\Psi}
   \>,
   \label{eq:heisy_2}
\end{equation}
which leads to the usual eigenvalue problem
\begin{equation}
   \sum_\beta \Bigl ( \meh{\Psi_\alpha}{\Psi_\beta}
   -
   E_\alpha \delta_{\alpha \beta} \Bigr )
   \,
   \dotp{\Psi_\beta}{\Psi} \ = \ 0 \>.
   \label{eq:heisy_3}
\end{equation}
We make the convention that Greek indices run over both the
discrete and continuum spectra, i.e. the sum $\sum_\alpha$ is in
fact a sum over the discrete, and an integral over the continuum.
When solving the above eigenvalue, one is usually forced to
discretize the continuum: one uses a discrete basis, which results
into a finite matrix, which can then be diagonalized using regular
linear algebra numerical packages.

We would like to avoid this approach. Therefore, rather than
calculating the ``entire'' spectrum of the hamiltonian at the same
time, we will attempt to calculate the spectrum of the hamiltonian
one state at a time, similar to the GFMC approach. In the same
spirit, we will attempt to calculate observables without explicit
knowledge of the actual wave functions: wave functions are not
observables, and we will concentrate on developing a formalism
allowing direct calculation of the expectation values of
operators.

In the CCE formalisms, one first introduces an ansatz for the
eigenstate $\ket{\Psi}$ of the hamiltonian
\begin{equation}
   \ket{\Psi} \ = \ e^{\s} \ \ket{\Phi} \>,
\label{eq:psi}
\end{equation}
where $\ket{\Phi}$ is the physical vacuum or reference state. The
ansatz for $\ket{\Psi}$ satisfies the normalization condition
\begin{equation}
   \langle \Psi | \Phi \rangle = 1
   \>.
\end{equation}
Here $\s$ denotes the many-body cluster-correlation operator:
\begin{equation}
   \s \ = \
   \sum_{n=1}^A \ 
       S_n \, \Ond{n}
   \>,
\end{equation}
where $\Ond{n}$ denotes an \nph{n}-configuration creation
operator. Mathematically speaking, the interpretation of a \nph{n}
configuration is that an element of a basis set spanning the
many-body Hilbert space.
For a given \nph{n} configuration, the correlation function for
$n$-nucleons, or the \nph{n} correlation coefficients $S_n$, is
simply the associated expansion coefficient.

We emphasize that the physical vacuum, $\ket{\Phi}$, is defined as
the minimal set of configurations obtained from the bare vacuum,
which obey a certain set of symmetries and boundary conditions.
Appropriate choices can be made for $\ket{\Phi}$ which will allow
us to independently construct both the ground state and excited
state spectrum of the hamiltonian, without changing the
single-energy spectrum. The cluster-correlation operator is a
rank-zero tensor operator and will leave unchanged both the
symmetry and boundary conditions of $\ket{\Phi}$. Therefore, the
information regarding a particular set of quantum numbers is
contained in the physical vacuum $\ket{\Phi}$; properties such as
orthogonality of the various eigenstates will be satisfied by
construction. In so doing, we depart from prior attempts to obtain
the excited state spectrum of the hamiltonian by means of an
equation of motion approach which relies on a prior calculation of
the ground state.

With these definitions, the Schr\"odinger equation for the
eigenstate $\ket{\Psi}$ of the hamiltonian corresponding to a
chosen symmetry,
\begin{equation}
   \h \ket{\Psi} = E \ket{\Psi} \>,
\label{eq:se}
\end{equation}
becomes
\begin{equation}
   e^{- \s} \h e^{\s} \ \ket{\Phi} \ = \ E \ \ket{\Phi} \>,
\end{equation}
or, in normal ordered form,
\begin{equation}
   \no{ e^{- \s} \h e^{\s} }{c} \ \ket{\Phi}
   \ = \ E \ \ket{\Phi} \>,
\label{eq:normal}
\end{equation}
where the subscript $c$ indicates the \emph{pure} creation part of
a normal ordered  operator. In principle, the resulting system of
integral equations must be solved for the \nph{n} correlation
amplitudes $S_n$ and the energy $E$. It is desirable, and in the
spirit of the approach we initiate in this paper, to design a way
of calculating $E$ without explicitly calculating (and storing!)
$S_n$.

\section{Observables}
\label{sec:obs}

Consider the expectation value of an arbitrary operator~$\Op$
\begin{equation}
   \bar o \ = \
   \frac{ \me{\Psi}{\Op}{\Psi} }{ \dotp{\Psi}{\Psi} }
   \>.
\end{equation}
We perform the following transformations
\begin{align*}
   \bar{o} \ = \ &
   \frac{ \me{\Phi}{e^{\s^\dag} \Op e^{\s}}{\Phi} }
        { \dotp{\Psi}{\Psi} }
   \\ \ = \ &
   \frac{ \me{\Phi}{e^{\s^\dag} e^{\s} \ e^{-\s} \Op e^{\s}}{\Phi} }
        { \dotp{\Psi}{\Psi} }
   \>.
\end{align*}
We can insert a complete set of \nph{n} configurations, and obtain
\begin{align*}
   \bar{o} \ = \ &
   \sum_n \
   \frac{ \me{\Phi}{e^{\s^\dag} e^{\s} \ \Ond{n}}{\Phi} }
        { \dotp{\Psi}{\Psi} } \
   \me{\Phi}{\On{n} \ e^{-\s} \Op e^{\s}}{\Phi}
   \\ \ = \ &
   \sum_n \
   \frac{ \me{\Psi}{ \Ond{n} }{\Psi} }
        { \dotp{\Psi}{\Psi} } \
   \me{\Phi}{\On{n} \ e^{-\s} \Op e^{\s}}{\Phi}
   \>,
\end{align*}
where in the last step we have used the fact that $\Ond{n}$ and
$\s$ commute. We introduce a new cluster-correlation operator,
$\tilde \s$, by its many-body decomposition, as
\begin{equation}
   \tilde \s \ = \
   \sum_n \
   \frac{ \me{\Psi}{ \Ond{n} }{\Psi} }
        { \dotp{\Psi}{\Psi} } \
   \On{n}
   \ = \
   \sum_n \ \tilde S_n \, \On{n}
   \>.
\end{equation}
Note that
\begin{equation}
   \tilde S_n
   \ = \
   \frac{ \me{\Psi}{ \Ond{n} }{\Psi} }
        { \dotp{\Psi}{\Psi} }
   \ = \
   \frac{ \me{\Psi}{ \On{n} }{\Psi} }
        { \dotp{\Psi}{\Psi} } \
   \>,
\end{equation}
and therefore the $\tilde S_n$ amplitudes are real. With this
definition, we can calculate the expectation value of the operator
$\Op$ as
\begin{equation}
   \bar o
   \ = \
   \me{\Phi}{\tilde \s \ e^{-\s} \Op e^{\s}}{\Phi}
   \ = \
   \me{\Phi}{e^{\s^\dag} \Op e^{- \s^\dag} \ \tilde \s^\dag }{\Phi}
   \>.
\label{eq:expect}
\end{equation}
Based on the (second) definition of the \nph{n} amplitudes
\begin{equation}
   \tilde S_n
   \ = \
   \me{\Phi}{\tilde \s \ e^{-\s} \On{n} e^{\s}}{\Phi}
   \>,
\end{equation}
we can determine $\tilde S_n$ via an iterative procedure.

Note that the correlated ground state $|\Psi\rangle$ is not a
translational-invariant wave function. Therefore, in practice one
must correct for the effects of the center-of-mass (CM) motion. In
our previous studies, a many-body expansion has been devised to
evaluate the corrections required by the calculation of
observables in the CM frame~\cite{ref:bmCM}. The accuracy of the
proposed procedure was tied to the success of a good separation of
the CM and relative coordinates degrees of freedom in the CCE
solution. This hypothesis was tested~\cite{ref:bm_II} by adding to
the Hamiltonian a purely CM piece, multiplied by a strength
parameter, and making sure that the binding energy is independent
of this parameter.

We review this issue now on fundamental grounds: In order to
produce translational-invariant many-body wave functions, one must
seek a simultaneous eigenvalue of both the hamiltonian and the
total momentum of the system:
\begin{align}
   \h \ \ket{\Psi} \ = \ E \ \ket{\Psi}
   \>,
   \\
   \pcm \ \ket{\Psi} \ = \ \vec P \ \ket{\Psi}
   \>,
\end{align}
or
\begin{align}
   e^{-\s} \h e^{\s} \ \ket{\Phi} \ = \ E \ \ket{\Phi}
   \>,
\label{eq:es}
   \\
   e^{-\s} \pcm e^{\s} \ \ket{\Phi} \ = \ \vec P \ \ket{\Phi}
   \>.
\label{eq:pcm}
\end{align}
A closer inspection of the above set of equations shows that this
not an over-determined system of equations: By definition, the
hamiltonian is the \emph{internal} hamiltonian of the many-body
system, where the CM kinetic energy has been removed. The
following commutators are identically zero:
\begin{align}
   [ \h , \, \pcm ]
   \ = \
   \bigl [  e^{-\s} \h  e^{\s} , \,  e^{-\s} \pcm  e^{\s}
   \bigr ]
   \ = \ 0
   \>.
\end{align}
One can replace Eq.~\eqref{eq:pcm} with a linear combination of
Eqs.~\eqref{eq:es} and \eqref{eq:pcm} :
\begin{align}
   e^{-\s} [ \h , \, \pcm ] e^{\s} \ \ket{\Phi} \ = \ 0
   \>.
\end{align}
Since the hamiltonian and the total momentum commute, it follows
that this equation is always satisfied by a solution of
Eq.~\eqref{eq:es}. Therefore, the \emph{dressed} hamiltonian, $
e^{-\s} \h e^{\s}$, does not modify the CM degrees of freedom
content of the physical vacuum, $\ket{\Phi}$.

In practice one has to truncate the CCE hierarchy of equations,
and an inadequate truncation may render the above discussion
inconsequential. However, the above statements remain valid for a
truncated version of the CCE, provided that a truncated
cluster-correlation operator $\s^{(N)}$ can be defined, and
$\s^{(N)}$ satisfies the identities
\begin{align}
   e^{-\s^{(N)}} \, e^{\s^{(N)}} \ = \ 1
   \>,
\end{align}
and
\begin{align}
   e^{- \s^{(N)}} \Op \ e^{\s^{(N)}} \ = \ &
   \Op + [ \Op, \s^{(N)} ]
   \\ \notag &
   + \frac{1}{2!} \bigl [ [ \Op, \s^{(N)} ], \s^{(N)} \bigr ]
   + \cdots
   \>.
\end{align}
An acceptable truncation of the CCE hierarchy is generated by the
prescription
\begin{equation}
   S_n = 0, \ n > N \>.
\end{equation}
The physical vacuum plays indeed a key role in the CCE formalism:
When done correctly, and even if $\ket{\Phi}$ breaks certain
symmetries of the desired solution, the CCE will leave the
center-of-mass part of the physical vacuum unchanged. Therefore,
one can perform corrections at the level of the physical vacuum,
and the further inclusion of the \nph{n} correlations will leave
these corrections unchanged. This is an important observation,
since we have the physical vacuum in closed form, and we would
rather not explicitly calculate the \nph{n} correlations. If
$\ket{\Phi}$ can be written as the product of a CM wave function
and a wave function of the system with respect to the
center-of-mass, then one can carry out the CM corrections via a
simple correction factor, just like in the shell model with
harmonic oscillator wave functions.

\section{$A=1$ : Single-particle quantum mechanics}
\label{sec:a1}

So far, our discussion has been quite general. We will
particularize now to the simplest possible case, namely the case
of a single-particle quantum mechanics problem. The CCE is a
many-body formalism, but the procedure is quite general, and we
can apply it to the $A=1$ case. In this context, we will continue
using the many-body language, and refer to ``configurations" and
``correlations", but solely because this is an intrinsic part of
the CCE formalism.

In the one-body case, the cluster-correlation operator is
particularly simple
\begin{equation}
   \s \ = \ 
   \s_1
   \>.
\label{eq:s_s1}
\end{equation}
Since the cluster-correlation operator contains only \nph{1}
correlations Eq.~\eqref{eq:psi} can be written as
\begin{equation}
   \ket{\Psi} \ = \ e^{\s} \ \ket{\Phi} \ = \ (\mathbf{1} + \s) \ \ket{\Phi}
   \>,
\label{eq:f1}
\end{equation}
where
no matter what symmetry we are interested in, there
is only one possible configuration in $\ket{\Phi}$. We introduce
the physical vacuum for the one-body case, as
\begin{equation}
   \ket{\Phi} \ = \ \ket{\phi_h} \>.
\end{equation}
For simplicity, we will consider that $\ket{\phi_h}$ is the only
state in the discrete spectrum satisfying the required symmetry.
An extension to the case when we have more than one discrete state
in the spectrum is straightforward and the relevant equations will
be relegated to Appendix~\ref{app:En}.

In order to derive the corresponding CCE equations we use the
identity (see also Appendix~\ref{app:gs_comm})
\begin{eqnarray}
   e^{- \s} \h e^{\s} & = &
   \h + [ \h, \s_1 ]
   + \frac{1}{2!} \bigl [ [ \h, \s_1 ], \s_1 \bigr ]
   \>.
\end{eqnarray}
Then, Eq.~(\ref{eq:normal}) is equivalent to the following set of
equations
\begin{eqnarray}
   E & = &
   \no{ \h }{0}
   + \no{ [ \h, \s_1 ] }{0}
   \>,
   \\
   0 & = &
   \no{ \h }{1}
   + \no{ [ \h, \s_1 ] }{1}
   + \frac{1}{2!} \no{ \bigl [ [ \h, \s_1 ], \s_1 \bigr ] }{1}
   \>.
\end{eqnarray}
These equations are solved for the energy $E$ corresponding to
$\ket{\Psi}$, and the correlations $S_1$.

Using Eqs.~\eqref{eq:s_s1} and \eqref{eq:f1} we can write the
solution of the Schr\"odinger equation as
\begin{equation}
   \ket{\Psi} \ = \ 
   \ket{\phi_n} + \dmom{p} \msa{p}{h} \ket{\phi_{\vec p}}
\label{eq:psi_eig}
   \>.
\end{equation}
We note that for the one-body problem, it is particularly simple
to find an interpretation or an independent check for the CCE
equations. Based on Eqs.~\eqref{eq:f1} and \eqref{eq:se}, we can
readily derive
\begin{align}
   E \ = \ &
   \meh{\phi_h}{\Psi}
\label{eq:s1_e}
   \>,
   \\
   E \, \msa{p}{h} \ = \ &
   \meh{\phi_p}{\Psi}
\label{eq:s1_s1}
   \>.
\end{align}

Using the commutators listed in Appendix~\ref{app:gs_comm}, the
CCE equations we need to solve become
\begin{align}
   E &\ = \ \meh{\phi_h}{\phi_h}
   + \dmom{p} \meh{\phi_h}{\phi_{\vec p}} \, \msa{p}{h}
   \>,
\label{eq:P_energy}
   \\
   0 & = \meh{\phi_{\vec p}}{\phi_h}
   \nonumber \\ &
   + \dmom{p_1} \meh{\phi_{\vec p}}{\phi_{\vec p_1}} \, \msa{p_1}{h}
   - \msa{p}{h} \, \meh{\phi_h}{\phi_h}
   \nonumber \\ &
   - \dmom{p_1}
     \msa{p}{h} \, \meh{\phi_h}{\phi_{\vec p_1}} \, \msa{p_1}{h}
   \>.
\label{eq:P_s1}
\end{align}
By inspection, the last equation gives
\begin{align}
   E \, \msa{p}{h}
   = &
   \meh{\phi_{\vec p}}{\phi_h}
   + \dmom{p_1} \meh{\phi_{\vec p}}{\phi_{\vec p_1}} \, \msa{p_1}{h}
   \>.
\label{eq:P_s1_p}
\end{align}
Equations~\eqref{eq:P_energy} and \eqref{eq:P_s1_p} involve
explicit integrals over correlations in momentum space, and we
will refer to this set of equations as the $P$-set. These
equations are entirely consistent with Eqs.~\eqref{eq:s1_e} and
\eqref{eq:s1_s1} above.

We note for completeness, that by formally introducing the
notations
\begin{align}
   v_{\vec p} \ = \ &
   \meh{\phi_{\vec p}}{\phi_h}
   \>,
   \\
   v_{\vec p}^T \ = \ &
   \meh{\phi_h}{\phi_{\vec p}}
   \>,
   \\
   A_{\vec p_1 \vec p_2} \ = \ &
   E \, \delta_{\vec p_1 \vec p_2}
   -
   \meh{\phi_{\vec p_1}}{\phi_{\vec p_2}}
   \>,
\end{align}
the solution of $P$-set of equations can be formally obtained as
\begin{align}
   S_1 \ = \ &
   [A]^{-1} \, v
   \>,
   \\
   E \ = \ &
   \meh{\phi_h}{\phi_h}
   \ + \ v^T \, \cdot \, [A]^{-1} \, v
   \>.
\end{align}
This implies, of course, that in order to get the energy, one has
again to discretize the continuum, and the solution is obtained by
inverting a matrix. The procedure is iterated until convergence is
achieved. We do not propose using this approach.

The above momentum space integrals can be avoided in favor of a
coordinate representation: We begin by introducing the
\emph{hole}-function
\begin{equation}
   \chi_h(\vec r) \ = \
   \dmom{p} \phi_{\vec p}(\vec r) \ \msa{p}{h}
   \>,
\label{eq:chi_def}
\end{equation}
or
\begin{equation}
   \ket{\chi_h} \ = \
   \dmom{p} \ket{\phi_{\vec p}} \ \msa{p}{h}
   \>.
\end{equation}
Note that in this simple one-body case, we have
\begin{equation}
   \ket{\chi_h} \ = \
   \ket{\Psi} - \ket{\phi_h}
\label{eq:chi_eig}
   \>.
\end{equation}
With this notation, we obtain the equivalent set of equations
\begin{align}
   E = & \meh{\phi_h}{\phi_h}
   + \meh{\phi_h}{\chi_h}
   \>,
\label{eq:energy}
   \\
   E \, \msa{p}{h} = &
   \meh{\phi_{\vec p}}{\phi_h}
   + \meh{\phi_{\vec p}}{\chi_h}
   \>,
\label{eq:s1_p}
\end{align}
where the $\msa{p}{h}$ correlations are interpreted as projections
of the \emph{hole}-function, $\ket{\chi_h}$, onto the continuum
set of states, i.e.
\begin{equation}
   \msa{p}{h} \ = \
   \dotp{\phi_{\vec p}}{\chi_h}
   \>.
\end{equation}
We can transform Eq.~\eqref{eq:s1_p} into an equation for the
\emph{hole}-function: we multiply by $\ket{\phi_{\vec p} }$ , and
integrate over the momentum variable. 
Using the closure relationship, Eq.~\eqref{eq:complete}, we obtain
\begin{align}
   E \, \ket{\chi_h} \ = \ &
   \h \, \ket{\phi_h}
   + \h \, \ket{\chi_h}
   \\ \notag &
   - \ket{\phi_h}
   \Bigl \{
      \meh{\phi_h}{\phi_h}
      + \meh{\phi_h}{\chi_h}
   \Bigr \}
   \>.
\end{align}
This equation is consistent with the eigenvalue equation for
$\ket{\Psi}$ :
\begin{align*}
   ( \h - E ) \, \ket{\chi_h}
   +
   ( \h - E ) \,
      \ket{\phi_h}
   \ = \
   ( \h - E ) \, \ket{\Psi}
   \ = \ 0
   \>,
\end{align*}
and can be used to evaluate $\ket{\chi_h}$. We have :
\begin{align}
   ( \h - E ) \, \ket{\chi_h}
   \ = \ - \,
   ( \h - E ) \,
      \ket{\phi_h}
   \>,
\label{eq:chi_h}
\end{align}
subject to the constraint that $\ket{\chi_h}$ is orthogonal to the
discrete basis states, i.e.
\begin{equation}
   \dotp{\phi_n}{\chi_h} \ = \ 0 \>.
\end{equation}
At this point we consider a new complete set of functions in
coordinate space, which satisfy the required boundary conditions.
Since these functions may overlap with the discrete states of the
original basis, we will remove this overlap by using the procedure
outlined in Eq.~\eqref{eq:overlap}. We will then expand
$\ket{\chi_h}$ in the new basis, and use Eq.~\eqref{eq:chi_h} to
determine the corresponding expansion coefficients. A similar
technique is successfully used in continuum RPA
theory~\cite{ref:crpa}. The approach is also similar to a method
introduced many years ago by Podolsky~\cite{ref:P28,ref:FF}, which
allows for the substitution of an integral over the continuum part
of the spectrum in terms of a bound-state eigenfunction of an
auxiliary hamiltonian. Podolsky's method was initially introduced
to treat dispersion in hydrogen atoms, and it has been recently
applied to the calculation of the electric polarizability of the
deuteron~\cite{ref:FP} with various realistic $N\!N$ potentials.

We will refer to the set of equations \eqref{eq:energy}
and \eqref{eq:chi_h}, as the $R$-set. The key element in this
representation is the fact that an explicit knowledge of the
continuum part of the basis states, $\ket{\phi_{\vec p}}$, is not
mandatory.


We turn our attention now to the calculation of observables: For a
one-body problem, $\tilde \s$ contains only \nph{1} configurations
\begin{equation}
   \tilde \s \ = \ {\mathbf 1} + \tilde \s_1
   \>.
\end{equation}
We determine $\tilde S_1$ as a solution of the equation
\begin{align}
   \mta{p}{h} \ = \ &
   \me{\Phi}{e^{-\s} \, \B{h} \A{p} \ e^{\s}}{\Phi}
   \\ \notag &
   +
   \dmom{p_1} \, \mta{p_1}{h} \,
   \me{\Phi}{\B{h} \A{p_1} \ e^{-\s} \, \B{h} \A{p} \ e^{\s}}{\Phi}
   \>,
\end{align}
and subsequently, observables will be calculated as
\begin{align}
   \bar o \ = \ &
   \me{\Phi}{e^{-\s} \Op e^{\s}}{\Phi}
   \\ \notag &
   +
   \dmom{p} \, \mta{p}{h} \,
   \me{\Phi}{\B{h} \A{p} \ e^{-\s} \Op e^{\s}}{\Phi}
   \>.
\end{align}
After working out the necessary commutators, we obtain
\begin{align}
   &
   \mta{p}{h} \ = \
   \msa{p}{h} \
   \Bigl [
      1
      -
      \dmom{p_1} \mta{p_1}{h} \msa{p_1}{h}
   \Bigr ]
   \>,
\label{eq:stil}
   \\ &
   \bar o \ = \
\label{eq:expect_o}
   \me{\phi_h}{\Op}{\Psi}
   \\ \notag & \qquad \quad
      + \dmom{p} \mta{p}{h} \Bigl [
      \me{\phi_{\vec p}}{\Op}{\Psi}
      -
      \me{\phi_h}{\Op}{\Psi}
      \msa{p}{h}
   \Bigr ]
   \>.
\end{align}
These equations involve integrals in momentum space, and
therefore, Eqs.~\eqref{eq:stil} and \eqref{eq:expect_o} represent
a natural complement to the $P$-set of equations discussed above.
In order to derive the corresponding coordinate representation, we
need to introduce a \emph{modified} \emph{hole}-function
\begin{equation}
   \tilde \chi_h(\vec r) \ = \
   \dmom{p} \phi_{\vec p}(\vec r) \ \mta{p}{h}
   \>,
\label{eq:tilchi_def}
\end{equation}
or
\begin{equation}
   \ket{\tilde \chi_h} \ = \
   \dmom{p} \ket{\phi_{\vec p}} \ \mta{p}{h}
   \>.
\end{equation}
Note that by construction we have $\dotp{\tilde \chi_h}{\phi_n} =
0$. With this notation, the above Eqs.~\eqref{eq:stil} and
\eqref{eq:expect_o} become
\begin{align}
   &
   \bra{\tilde \chi_h} \ = \
   \bra{\chi_h} \
   \Bigl [
      1
      -
      \frac{1}{E} \, \meh{\tilde \chi_h}{\Psi}
   \Bigr ]
\label{eq:tils1}
   \>,
   \\ &
   \bar o \ = \
   \me{\phi_h}{\Op}{\Psi}
   + \me{\tilde \chi_h}{\Op}{\Psi}
   - \frac{1}{E} \, \meh{\tilde \chi_h}{\Psi}
     \me{\phi_h}{\Op}{\Psi}
   \label{eq:baro}
   \>.
\end{align}
These equations allow for the calculation of the observable $\bar
o$ entirely in coordinate space.

An interesting expression for
the expectation value of an arbitrary operator 
is obtained by combining Eqs.~\eqref{eq:tils1} and
\eqref{eq:baro}. We obtain
\begin{equation}
   \bar o \ = \
   \me{\Psi}{\Op}{\Psi} \
   \Bigl [
      1
      -
      \frac{1}{E} \, \meh{\tilde \chi_h}{\Psi}
   \Bigr ]
   \>.
\label{eq:baro_sub}
\end{equation}
This last equation is a direct consequence of the fact that CCE
does not automatically generate a normalized wave function
$\ket{\Psi}$. Of course, for the simple case we consider here, a
brute force normalization of $\ket{\Psi}$ is always possible, but
such a procedure will be impossible for $A>1$. Our
prescription~\eqref{eq:expect_o} for calculating the expectation
value of $\Op$ provides an implicit normalization for the purpose
of calculating the observable~$\bar o$, by means of a simple
substraction from the expectation value calculated using the
unnormalized wave function $\ket{\Psi}$, as shown in
Eq.~\eqref{eq:baro_sub}.

\section{Iterative solutions}
\label{sec:iterative}

Let us consider the $P$-set of equations, Eqs.~\eqref{eq:P_energy}
and \eqref{eq:P_s1_p}. An iterative solution for the \nph{1}
correlations, $\msa{p}{h}$, can be obtained by iterating
Eq.~\eqref{eq:P_s1_p}. We have
\begin{align}
   \msa{p_1}{h} \ = \ &
   \frac{1}{E} \, \meh{\phi_{\vec p_1}}{\phi_h}
   +
   \frac{1}{E^2} \, \meh{\phi_{\vec p_1}}{\phi_{\vec p_2}}
   \meh{\phi_{\vec p_2}}{\phi_h}
\notag      \\ &
   +
   \cdots
   +
   \frac{1}{E^{n-1}} \, \meh{\phi_{\vec p_1}}{\phi_{\vec p_2}}
   \cdots
   \meh{\phi_{\vec p_{n}}}{\phi_h}
   +
   \cdots
\label{eq:s1_iter}
   \>.
\end{align}
Implicit integration over all repeated continuum indices is
assumed. By combining Eqs.~\eqref{eq:s1_iter} and
\eqref{eq:P_energy}, we obtain a polynomial equation for the
energy of the desired state
\begin{align}
   E \ = \ &
\label{eq:E_iter}
   \meh{\phi_h}{\phi_h}
   + \meh{\phi_h}{\phi_{\vec p_1}}
   \Bigl [ \
   \frac{1}{E} \, \meh{\phi_{\vec p_1}}{\phi_h}
   \\ \notag &
   +
   \frac{1}{E^2} \, \meh{\phi_{\vec p_1}}{\phi_{\vec p_2}}
   \meh{\phi_{\vec p_2}}{\phi_h}
   \\ \notag &
   +
   \cdots
   +
   \frac{1}{E^{n-1}} \, \meh{\phi_{\vec p_1}}{\phi_{\vec p_2}}
   \cdots
   \meh{\phi_{\vec p_{n}}}{\phi_h}
   +
   \cdots
   \Bigr ]
   \>.
\end{align}
The correlations $\msa{p}{h}$ do not explicitly enter the last
equation.

The above equations are not suitable for a Monte Carlo approach to
obtaining the energy $E$, since the energy itself appears as a
inverse power factor in both Eqs.~\eqref{eq:s1_iter} and
\eqref{eq:E_iter}. This situation is an artifact of the one-body
problem we consider here, and can be traced back to the simple
substitution made in deriving Eq.~\eqref{eq:P_s1_p} from
~\eqref{eq:P_s1}. For the actual many-body ($A>1$) problem, the
$S_n$ equations will also feature an energy denominator, but $E$
will be replaced by the \nph{n}-energy, $\varepsilon_n= \{
\epsilon_{\vec p_1} - \epsilon_h, \ldots \}$. The existence of an
energy gap between the discrete and continuum spectrum will
provide a \emph{small} parameter for this expansion.

Nevertheless, Eq.~\eqref{eq:E_iter} deserves a second look: We use
the closure relation~\eqref{eq:complete} to eliminate the
integrals over the continuum basis states. We obtain
\begin{align}
   E \ = \ &
   \meh{\phi_h}{\phi_h}
   + \frac{1}{E}
   \Bigl \{ \me{\phi_h}{\h^2}{\phi_h} - [\me{\phi_h}{\h}{\phi_h}]^2
   \Bigr \}
   \notag \\ &
   + \frac{1}{E^2}
   \Bigl \{ \me{\phi_h}{\h^3}{\phi_h}
      - 2 \me{\phi_h}{\h^2}{\phi_h} \me{\phi_h}{\h}{\phi_h}
   \notag \\ & \qquad \qquad
      + [\me{\phi_h}{\h}{\phi_h}]^3
   \Bigr \}
   +
   \cdots
\end{align}
This equation can be used in practice to calculate $E$ for $A=1$,
by considering successive truncations of the above series, and
finding the root of the resulting polynomial equation in $E$.

An iterative scheme is also possible for calculating observables.
The observable $\bar o$ can be obtained by employing
Eq.~\eqref{eq:tils1} and a repeated substitution of $\bra{\tilde
\chi_h}$ in Eq.~\eqref{eq:baro_sub}. We have
\begin{align}
   \bar o \ = \
   \me{\Psi}{\Op}{\Psi} \
      \sum_{k=0}^{\infty} \ \frac{(-1)^k}{E^k} \ [\meh{\chi_h}{\Psi}]^k
   \>.
\end{align}

To conclude, let us describe another traditional strategy for
obtaining the numerical solution of Eq.~\eqref{eq:P_s1_p}, which
also relies on an iterative procedure: We begin by making an
initial guess for the \nph{1} correlations:
\begin{equation}
   \msa{p}{h}^{0} = - \,
   \frac{ \meh{\phi_{\vec p}}{\phi_h} }
        { \varepsilon_{\vec p} - \varepsilon_h }
   \>,
\end{equation}
where we use the notation $\varepsilon_\alpha =
\meh{\phi_\alpha}{\phi_\alpha}$, the diagonal matrix elements of
the hamiltonian. Then, we improve on this initial guess, by
finding a first-order correction to $\msa{p}{h}^{0}$ in the
standard fashion:
\begin{equation}
   \msa{p}{h}^{1} = \msa{p}{h}^{0} + \epsilon_{\vec p h}
   \>.
\end{equation}
This procedure is iterated until convergence is achieved. The
correction $\epsilon_{\vec p h}$ is the solution of the equation
\begin{align}
   &
   E^{0} \, \epsilon_{\vec p h}
   - \dmom{p_1} \meh{\phi_{\vec p}}{\phi_{\vec p_1}} \, \epsilon_{\vec p_1 h}
\label{eq:to_MC}
   \\ \notag &
   + \msa{p}{h}^{0} \,
     \dmom{p_1} \meh{\phi_h}{\phi_{\vec p_1}} \, \epsilon_{\vec p_1 h}
   \\ \notag & \qquad
   =
   \meh{\phi_{\vec p}}{\phi_h}
   + \dmom{p_1} \meh{\phi_{\vec p}}{\phi_{\vec p_1}} \, \msa{p_1}{h}^{0}
   - E^{0} \, \msa{p}{h}^{0}
   \>,
\end{align}
where $E^{0} \equiv E[\msa{p}{h}^0]$. Of course, this procedure
also relies on discretizing the continuum, and involves a matrix
inversion required for inferring the correction $\epsilon_{\vec p
h}$ as solution of the above system of linear equations. However,
given a particular hamiltonian $\h$, it is important to ask the
question wether this procedure is stable and converges to the
correct solution. If this is indeed the case, then one can show
that the solution of the CCE equations can be obtained as the sum
of an infinite Neumann-like series, and one can envision
attempting a numerical summation of this series by means of a
nonlinear Monte Carlo approach. The detailed discussion of this
approach is beyond the scope of the present paper.

\section{Numerical Digressions}
\label{sec:numerics}

At this point, a numerical study of some of the above formal
statements might be helpful to illustrate our point of view, and
the case of a simple single-particle quantum mechanics problem
provides interesting insight in the nature of the approximations.
Let us consider a spherically symmetric Wood-Saxon (WS) potential
well
\begin{equation}
   V(r) \ = \ - V_0 / \Bigl \{ 1 \ + \ \exp[ (r - R) / a] \Bigr \}
   \>,
\end{equation}
and address the problem of finding the bound-state spectrum of the
associated one-body hamiltonian. The choice of a WS potential is
not entirely accidental: the WS potential may play a special role
in our future many-body studies, as it mimics well the mean-field
nuclear potential, and accurate numerical methods for calculating
the bound state energies and wave functions are readily available
in the literature. As such, we may use single-particle WS wave
functions to construct good variational representations of the
physical vacuum in the CCE.

\begin{table*}[t]
\caption{\label{tab:erg}Energies [MeV] of WS bound states:
calculated values using the shooting method (exact), direct matrix
diagonalization in a $\mathcal{HO}$ basis (Heisenberg), physical
vacuum energy ($E_h$), and CCE using a pure $\mathcal{HO}$ basis
($\exp(\s)$: $\mathcal{HO}$), and a combination of a
$\mathcal{HO}$ and $\mathcal{L}$ basis ($\exp(\s)$:
$\mathcal{HO}$), respectively. Where $\mathcal{HO}$ basis are
concerned, results are reported for three values of the oscillator
parameter $b$, centered around the optimum $b$ value for the
variational ground state energy, $b=1.4$ fm .}
\begin{ruledtabular}
\begin{tabular}{ccrrrrr}
state & & Heisenberg &  $E_h$ & $\exp(\s)$: $\mathcal{HO}$ &
$\exp(\s)$: $\mathcal{HO}+\mathcal{L}$ & exact
\\
 & $b = 1.3 \ \mathrm{fm}$ & & & & &
\\
$[\ell=0, \, n=1]$ &&  -71.604115 & -71.154641 & -71.604115 & -71.604093 & \\
$[\ell=0, \, n=2]$ &&  -27.101582 & -26.341419 & -27.487394 & -27.101544 & \\
$[\ell=1, \, n=1]$ &&  -50.549663 & -50.189162 & -50.549663 & -50.549664 & \\
$[\ell=1, \, n=2]$ &&   -7.974057 &  -4.649952 &  -8.272408 &  -7.974058 & \\
$[\ell=2, \, n=1]$ &&  -28.555288 & -28.070778 & -28.555288 & -28.555288 & \\
$[\ell=3, \, n=1]$ &&   -6.890894 &  -5.580267 &  -6.890894 &  -6.890894 & \\
 & $b = 1.4 \ \mathrm{fm}$ & & & & &
\\
$[\ell=0, \, n=1]$ &&  -71.604123 & -71.561383 & -71.604123 & -71.604093 &  -71.604071 \\
$[\ell=0, \, n=2]$ &&  -27.101595 & -26.875013 & -27.102023 & -27.101544 &  -27.101509 \\
$[\ell=1, \, n=1]$ &&  -50.549663 & -50.462831 & -50.549663 & -50.549664 &  -50.549664 \\
$[\ell=1, \, n=2]$ &&   -7.974057 &  -6.549628 &  -8.004096 &  -7.974058 &   -7.973692 \\
$[\ell=2, \, n=1]$ &&  -28.555288 & -28.368573 & -28.555288 & -28.555288 &  -28.555265 \\
$[\ell=3, \, n=1]$ &&   -6.890894 &  -6.352976 &  -6.890894 &  -6.890894 &   -6.890848
\\
 & $b = 1.5 \ \mathrm{fm}$ & & & & &
\\
$[\ell=0, \, n=1]$ &&  -71.604133 & -71.241552 & -71.604133 & -71.604093 & \\
$[\ell=0, \, n=2]$ &&  -27.101611 & -26.431344 & -27.393398 & -27.101544 & \\
$[\ell=1, \, n=1]$ &&  -50.549663 & -49.640614 & -50.549663 & -50.549664 & \\
$[\ell=1, \, n=2]$ &&   -7.974057 &  -7.578719 &  -8.724761 &  -7.974058 & \\
$[\ell=2, \, n=1]$ &&  -28.555288 & -27.396344 & -28.555288 & -28.555288 & \\
$[\ell=3, \, n=1]$ &&   -6.890894 &  -5.845401 &  -6.890894 &  -6.890894 & \\
\end{tabular}
\end{ruledtabular}
\end{table*}

\begin{figure*}[t]
   \includegraphics[width=\textwidth]{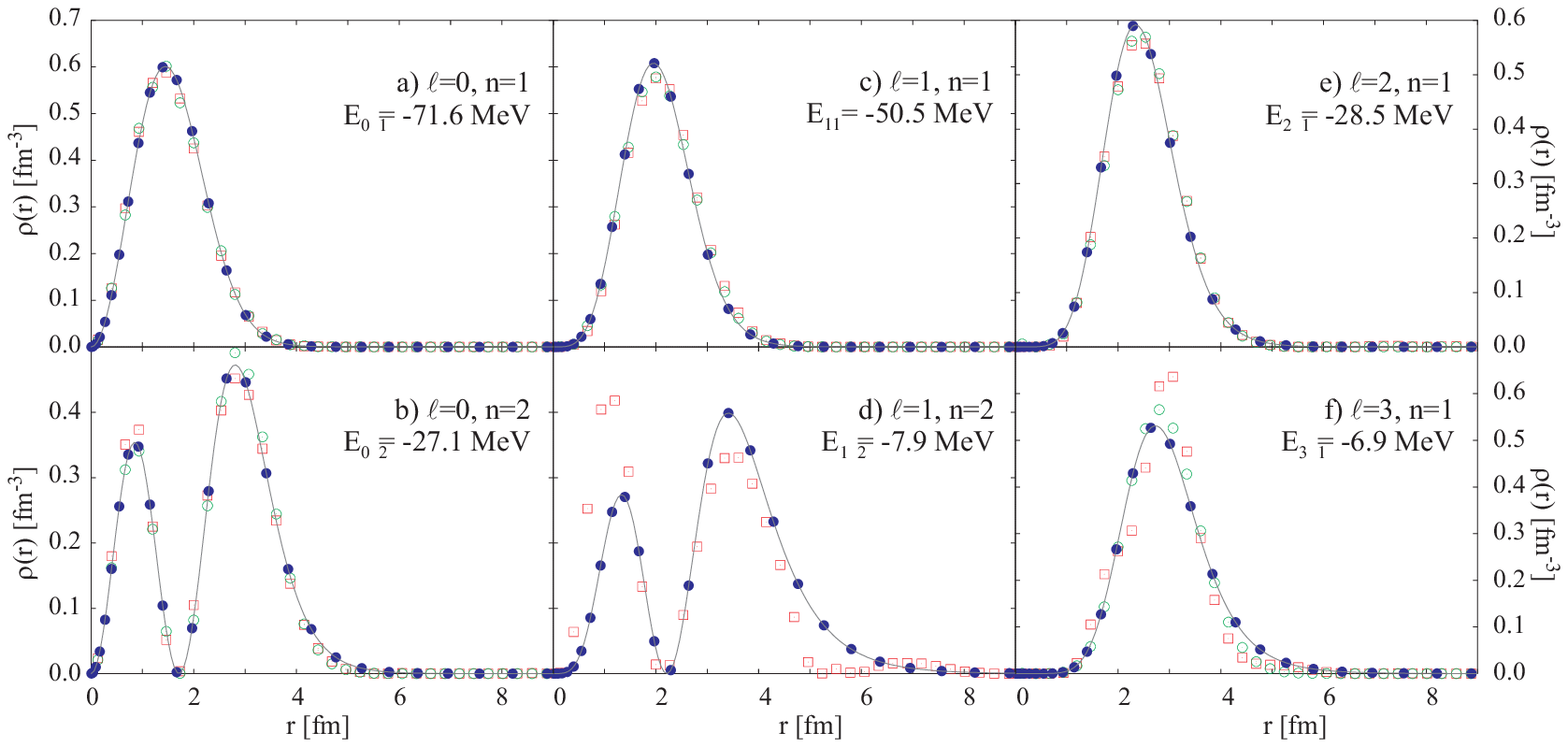}
   \caption{\label{fig:densities}(Color online) Density probabilities $\rho(r)$ 
of WS bound states, derived using the ``exact'' shooting method
(line), Heisenberg representation in a $\mathcal{HO}$ basis
(boxes), CCE using a single $\mathcal{HO}$ basis (empty circles),
and CCE using a combination of a $\mathcal{HO}$ and $\mathcal{L}$
basis (filled circles). Where appropriate, the values of the
employed parameters were $b=1.4$ fm, and $\gamma=3$ fm$^{-1}$. All
discrete basis featured $N_g=68$ functions, corresponding to the
identical number of collocation point for the Gauss-Hermite or
Gauss-Laguerre quadrature method, respectively.}
\end{figure*}

We consider a WS potential well with depth $V_0 = 100$ MeV, radius
$R=3$ fm, and diffuseness parameter $a=0.6$ fm. In the following
we focuss on the calculation of the bound-state energies, and the
corresponding probability densities. Numerical results are
collected in Table~\ref{tab:erg} and
Figs.~\ref{fig:densities}$a\div f$.

We begin by employing two textbook~\cite{ref:SSL} methods for
finding the bound-state eigenvalues of the hamiltonian. Firstly,
in order to establish a baseline for our comparison, we employ a
shooting technique and solve the radial Schr\"odinger equation on
a grid using the Numerov method. The radial wave function
calculation begins at the origin 
and the numerical solution is stepped out for a fixed energy $E$,
to $r_{\mathrm{max}}=12$ fm in $2^{12}\times 10^3$ steps. In the
presence of an eigenvalue, $E_{n\ell}$, the
wave function flips between large negative and 
positive values for small variations in the energy $E$. The energy
$E$ is fine tuned in order to capture this transition. In this
limit $E$ tends to $E_{n\ell}$, the true eigenvalue of the
Schr\"odinger equation. For $E=E_{n\ell}$, the numerical wave
function becomes normalizable over the radial interval
$[0,r_{\mathrm{max}}]$, and we can compute the associated density
probability on the interval. These eigenvalues and densities will
be referred to as ``exact'' for the purposes of our numerical
comparison.

Secondly, we obtain the solution of the Schr\"odinger equation in
a harmonic oscillator representation, by solving the eigenvalue
problem illustrated in Eq.~\eqref{eq:heisy_3} . In this
(Heisenberg) picture, the continuum part of the hamiltonian
spectrum is inherently discretized. The harmonic oscillator
($\mathcal{HO}$) wave functions depend on the choice of the
oscillator parameter $b$, and this in turn introduces a dependence
of the calculated spectrum on the oscillator parameter $b$. The
corresponding eigenfunctions are linear superpositions of
$\mathcal{HO}$ functions, which partially compensates for the fact
that the $\mathcal{HO}$ functions do not exhibit the corrected
asymptotic behavior demanded by the Wood-Saxon potential. We
consider a number of $N_g=68$ functions in our $\mathcal{HO}$
expansion, to correspond to the $N_g=68$ collocation points used
in our Gauss-Hermite quadrature scheme. Test runs for $N_g=52$
show that both the bound-state eigenvalues and eigenvectors of
this hamiltonian have largely saturated with $N_g$, and the
subsequent improvements are only minute with increasing $N_g>52$.
The eigenvalues are very close to the expected exact results. The
calculated densities for the lower energy states are reasonably
close to the exact result: the more bound the state is, the better
the density agreement. For the shallow ($E> - \,10$ MeV)
bound-states however, the calculated densities compare poorly with
the exact results: the model space is simply too small to allow
for a good representation of the wave function, even though the
eigenvalue calculation has arguably converged already for this
space size.

We turn our attention now to the CCE solution. In
Appendix~\ref{app:cce_a1} we list the actual equations we need to
solve. In all our CCE calculations the physical vacuum
$\ket{\Phi_{n\ell}}$ is represented by the corresponding
$\mathcal{HO}_{n\ell}$ wave function. The discrete spectrum of the
hamiltonian is then determined \emph{one state at a time}, by
solving the $R$-set of equations described above. In this approach
one may hope to account \emph{exactly} for the existence of the
continuum part of the spectrum, by introducing the hole-functions
$\ket{\chi_{n\ell}}$ and effectively mapping this continuum onto
the bound-state part of the spectrum.

We consider here two qualitatively different scenarios: The first
scenario is identical in spirit with the approach used in the
recent implementations of the CCE~\cite{ref:bm_I,ref:bm_II} for
spin-isospin shell saturated nuclei: we adopt a \emph{one-basis}
realization of the CCE, where we use the remaining part of the
$\mathcal{HO}$ basis (after eliminating those state used to
represent the physical vacuum, and possible all other lower energy
states of the same symmetry), to construct the desired
hole-function $\ket{\chi_{n\ell}}$. The functions in this set are
definitely orthogonal to each other, but they do not form a
complete basis in the Hilbert space. However, the condition
$\dotp{\phi_{n\ell}}{\chi_{n\ell}}=0$ must be satisfied, and
indeed the above set of functions is complete in the subspace of
all possible $\ket{\chi_{n\ell}}$. However, a simple manipulation
of the equations reveals that introducing the hole-function
formalism in this picture does not amount to any new information.
We are working within the confines of a $\mathcal{HO}$ potential,
and no structure outside this space can be built into the desired
$\ket{\Psi_{n\ell}}$ eigenfunction of the hamiltonian. We have
effectively discretized the continuum, and the accuracy of the
calculated eigenvalues and densities is generally close to the
results in the Heisenberg picture.

The ground state $[\ell=0,\, n=1]$ has a variational minimum $E_h$
for a value $b=1.4$ of the oscillator parameter, and the CCE
result in this representation is more sensitive to the choice of
$b$ than the previous result in the Heisenberg representation.
This reflects the importance of a good choice for the physical
vacuum for the CCE: a true variational calculation is necessary to
establish the best possible choice of $b$. In reviewing the
results, we note on the plus side that this version of the CCE
fixes some of the erratic behavior of the density for the
$[\ell=3,\, n=1]$ state. Still, we notice that too much strength
is pushed into the interior region, as the falloff of the tail is
too sharp. On the other hand, the CCE misses completely the second
state for $\ell=1$. From table~\ref{tab:erg} we see that already
at the physical vacuum level, the $\mathcal{HO}$ representation is
unable to provide an acceptable description of the $[\ell=1,\,
n=2]$ state: the vacuum energy $E_{h\equiv\{12\}}$ does not have a
minimum 
for $b \in [1.3,1.5]$, and the subsequent calculation of the
hole-function as a linear superposition of $\mathcal{HO}$
functions is unable to remedy this deficiency. Once again, the
wrong asymptotic behavior of the $\mathcal{HO}$ functions results
into a inadequate representation of the continuum, and thus the
serious shortcomings of the CCE in this representation.

In order to establish that this is indeed the case, we consider a
second (\emph{two-basis}) scenario in the CCE. We still represent
the physical vacuum as a simple $\mathcal{HO}_{n\ell}$ wave
function, but we expand the hole-functions using Laguerre
($\mathcal{L}$) functions as our \emph{second} complete basis.
Unlike the $\mathcal{HO}$ functions which falloff like $\exp[-r^2
/ (2 b^2) ]$, the asymptotic behavior of the $\mathcal{L}$
functions has the form $\exp(-\gamma r)$, similarly to the
bound-state coulomb functions. Properties of the $\mathcal{L}$
functions are summarized in Appendix~\ref{app:wfcn}, and for the
purpose of this two-set representation of the CCE, we are using a
Gauss-Laguerre quadrature scheme with $N_g=68$ collocation points.

It is interesting to remark that a CCE calculation done entirely
in $\mathcal{L}$ space (similar to the CCE calculation in the
$\mathcal{HO}$ space described above), results in good eigenvalues
for values of $\gamma=3 \div 4$, but fails to generate reasonable
densities: if the $\mathcal{HO}$ functions were falling off too
quickly with $r$, the $\mathcal{L}$ functions are spread out over
large distances and have little usable strength in the interior
region which can be used to build the density. When the two basis
are combined, however, the CCE results in the two-basis
representation literally collapse onto the exact results, for both
the bound state energies and densities. The numerical results show
very little sensitivity with the value of the
oscillator parameter $b$, for a fixed value of 
$\gamma = 3\, \mathrm{fm}^{-1}$: numerical differences for the
energies calculated for various values of $b$ appear beyond the
six significant digits presented here. The fact that we are
working with a \emph{finite} number ($N_g=68$) of basis states is
evident as the CCE result is not the same as the exact results,
but the error is less than 0.005\% .

The CCE calculation has no problem producing accurate energies
\emph{and} densities for the shallow bound states ($E>- \, 10$
MeV),
and the CCE successfully corrects for the lack of a viable
$\mathcal{HO}$ representation in the case of the physical vacuum
for the $[\ell=1,\, n=2]$ state. We would argue that a good
physical vacuum representation is still important, but a good
representation of the continuum is indeed crucial for an accurate
calculation of the bound-states spectrum of the hamiltonian.

\begin{table}[t]
\caption{\label{tab:norm}Restoration of normalization of WS
eigenfunctions in the CCE: using a pure $\mathcal{HO}$ basis, and
a combination of a $\mathcal{HO}$ and $\mathcal{L}$ basis,
respectively. All calculations correspond to the values of the
parameters $b=1.4$ fm, and $\gamma=3$ fm$^{-1}$, respectively.}
\begin{ruledtabular}
\begin{tabular}{crrrr}
state & $\dotp{\Psi}{\Psi}_{\mathcal{HO}}$ & $\rho_{\mathcal{HO}}$
      & $\dotp{\Psi}{\Psi}_{\mathcal{HO+L}}$ & $\rho_{\mathcal{HO+L}}$
\\
\\
$[\ell=0, \, n=1]$ &  1.000979 & 1.0005641 & 1.000415 & 1.000000 \\
$[\ell=0, \, n=2]$ &  1.000011 & 0.9954375 & 1.004602 & 1.000000 \\
$[\ell=1, \, n=1]$ &  1.000157 & 0.9988504 & 1.001309 & 1.000000 \\
$[\ell=1, \, n=2]$ &  1.006852 & 0.9323478 & 1.081757 & 1.000000 \\
$[\ell=2, \, n=1]$ &  1.001492 & 0.9985834 & 1.002913 & 1.000000 \\
$[\ell=3, \, n=1]$ &  1.000184 & 0.9823059 & 1.018200 & 1.000000 \\
\end{tabular}
\end{ruledtabular}
\end{table}

We conclude this analysis by considering the normalization issues
we have alluded to at the end of section~\ref{sec:a1}: we remind
the reader that the CCE wave function is not normalized in the
usual sense, but by requiring $\dotp{\Phi}{\Psi}=1$. Therefore,
the observables' calculation must implicitly ``correct'' for this
artifact, and ``restore'' the wave function normalization, see
Eq.~\eqref{eq:baro_sub}. In table~\ref{tab:norm}, we collect the
results regarding the normalization of the solution $\ket{\Psi}$,
and the normalization of the probability density $\rho(\vec r)$:
for an exact solution of the Schr\"odinger equation, the density
should come out automatically normalized to unity, even though
$\ket{\Psi}$ is not. Conversely, the departure from unity in the
$\rho(\vec r)$ normalization, is a test for the quality of the CCE
solution. Therefore, it comes to no surprise that the largest
discrepancies arise in the normalization of the $[\ell=3,\, n=1]$
and, particularly, $[\ell=1,\, n=2]$ states, in the one-basis
$\mathcal{HO}$ representation. The two-basis representation of the
CCE on the other hand, performs very well: the density
normalization is correct to 11 significant figures.

\section{Discussion and Outlook}
\label{sec:concl}

In this paper we have discussed the fundamental aspects of a new
formulation of the coupled-cluster expansion in the continuum, and
we have illustrated this approach by considering the simplest
($A=1$) dimensional realization of the many-body problem. For this
simple scenario we have derived CCE equations which can be solved
either in momentum or coordinate space. We have discussed the
calculation of the eigenvalue spectrum of the hamiltonian, as well
as the calculation of expectation values of arbitrary operators.

The momentum space approach seems to be the most promising one for
future efficient algorithms, with the drawback that an explicit
knowledge of the continuum basis states appears to be mandatory:
One can use Eqs.~\eqref{eq:P_energy} and \eqref{eq:P_s1} to design
an iterative, albeit nonlinear, approach for the direct
calculation of the energy, without an explicit reference to the
$S_1$ correlations. This may allow for a solution based on a
nonlinear Monte Carlo approach, contingent upon the favorable
resolution of serious concerns related to the numerical
convergence and stability, and the feasibility of a practical
numerical implementation. This discussion is tied in with the
convergence and stability issues of the iterative method outlined
in Section~\ref{sec:iterative}, following Eq.~\eqref{eq:to_MC}.
Work is currently ongoing to solve the technical issues of this
numerical approach.

The coordinate space approach is also promising, and likely to
offer immediate results. It offers the familiar perspective of a
coordinate space implementation, with the added advantage of a
lossless account of the continuum part of the single-energy
spectrum. This is realized via a small set of
\emph{hole}-functions, labelled after the states which are
occupied in the physical vacuum. However, a Monte Carlo approach
to solving this set of equations seems difficult to envision at
this time: The calculation of the \emph{hole}-functions employs an
expansion in a known (and hopefully finite) basis of functions,
with the expansion coefficients obtained as the numerical solution
of a system of linear equations derived from Eq.~\eqref{eq:chi_h}.

The stage is now set: We know that in the one-body case, we can
derive consistent CCE equations. We will derive next the CCE
equations for the deuteron ($A=2$) problem. In the CCE formalism,
one has an intermediate level of approximation short of the exact
solution, namely the $\s_1$ and $\s_2$ levels, respectively. We
will solve these equations using the realistic Argonne v$_{18}$
potential~\cite{ref:av18}, compare the approximate and exact CCE
solutions, and compare with results obtained using other exact
methods. In the end, the deuteron problem is extremely simple
outside the CCE formalism, since it can be formulated as an
one-body problem in terms of the relative coordinate of the system
of two nucleons.


\begin{acknowledgments}

B.M. would like to thank Jochen Heisenberg, Fritz Coester, and
John Dawson for their continued support and encouragement. The
author gratefully acknowledges useful discussions with Ben Gibson,
Jim Friar, Gerry Jungman, and Murray Peshkin. B.M. thanks Jim
Friar for pointing out the parallel between the hole-function
approach to CCE, and Podolsky's method.

\end{acknowledgments}


\appendix

\section{($A=1$) commutators}
\label{app:gs_comm}

We find useful to write the following anti-commutation properties
\begin{align}
    \{ \C{\vec p_1}, \Ad{p_2} \} & = \delta_{\vec p_1 \vec p_2} \>,
    \qquad
    \{ \C{\alpha}, \Bd{m_2} \} \ = \ 0 \>,
    \\
    \{ \Cd{\alpha}, \Ad{p_2} \} & = 0 \>,
    \qquad
    \{ \Cd{n_1}, \Bd{m_2} \} \ = \ \delta_{n_1 m_2} \>.
\end{align}
We list now the results used in deriving the CCE equations for our
\emph{toy} single-particle quantum mechanics problem:
\begin{itemize}

\item $\Cd{\alpha} \C{\beta}$
\begin{align*}
   &
   \Cd{\alpha} \C{\beta}
   \ = \
   ( \Ad{p_\alpha} + \B{n_\alpha} )
   ( \A{p_\beta} + \Bd{n_\beta} )
   \>.
\end{align*}
We obtain the relevant expectation values
\begin{gather}
   \bra{\Phi} \Cd{\alpha} \C{\beta} \ket{\Phi}
   \ = \ \delta_{n_\alpha n_\beta} \>,
   \\
   \bra{\Phi} \B{n_b} \A{p_a} \ \Cd{\alpha} \C{\beta} \ket{\Phi}
   \ = \ \delta_{\vec p_\alpha \vec p_a} \delta_{n_\beta n_b}
   \>.
\end{gather}

\item $[\h, \s_1]$
\begin{align*}
   &
   \bigl [ \Cd{\alpha} \C{\beta}, \, \Ad{p_1} \Bd{n_1} \bigr ]
   \\
   & =
   \Cd{\alpha} \CAdBd{\beta}{p_1}{n_1}
   + \CdAdBd{\alpha}{p_1}{n_1} \C{\beta}
   \\
   & = \delta_{\vec p_\beta \vec p_1} \Cd{\alpha} \Bd{n_1}
       - \, \delta_{n_\alpha n_1} \Ad{p_1} \C{\beta}
   \>.
\end{align*}
Note that
\begin{align*}
   &
   \bigl [ \Cd{\alpha} \C{\beta}, \, \Ad{p_1} \Bd{n_1} \bigr ] \ket{\Phi}
   \\ & =
   \Bigl (
          \delta_{\vec p_\beta \vec p_1} \Cd{\alpha} \Bd{n_1}
     - \, \delta_{n_\alpha n_1} \Ad{p_1} \Bd{n_\beta} \Bigr ) \ \ket{\Phi}
   \>.
\end{align*}
Therefore, the relevant expectation values are
\begin{align}
   &
   \bra{\Phi} \bigl [ , \, \bigr ] \ket{\Phi}
   \ \doteq \
   \delta_{\vec p_\beta \vec p_1} \delta_{n_\alpha n_1}
   \>,
   \\
   &
   \bra{\Phi} \B{n_b} \A{p_a} \ \bigl [ , \, \bigr ] \ket{\Phi}
   \\ \notag
   & \doteq
   \delta_{\vec p_\beta \vec p_1} \delta_{\vec p_\alpha \vec p_a} \delta_{n_1 n_a}
   - \, \delta_{n_\alpha n_1} \delta_{\vec p_1 \vec p_a} \delta_{n_\beta n_b}
   \>.
\end{align}

\item $\bigl [ [\h, \s_1], \, \s_1 \bigr ]$
\begin{align*}
   &
   \Bigl [
         \bigl [ \Cd{\alpha} \C{\beta}, \, \Ad{p_1} \Bd{n_1} \bigr ] , \,
         \Ad{p_2} \Bd{n_2} \Bigr ]
   \\
   & = \delta_{\vec p_\beta \vec p_1} \CdAdBd{\alpha}{p_2}{n_2} \Bd{n_1}
       - \, \delta_{n_\alpha n_1} \Ad{p_1} \CAdBd{\beta}{p_2}{n_2}
   \\
   & = - \, \delta_{n_\alpha n_2} \delta_{\vec p_\beta \vec p_1} \Ad{p_2} \Bd{n_1}
       - \, \delta_{n_\alpha n_1} \delta_{\vec p_\beta \vec p_2} \Ad{p_1} \Bd{n_2}
   \>.
\end{align*}
We have
\begin{align}
   &
   \bra{\Phi} \bigl [ [ , \, ], \, \bigr ] \ket{\Phi}
   \ = \ 0
   \>,
   \\
   &
   \bra{\Phi} \B{n_b} \A{p_a} \ \bigl [ [ , \, ], \, \bigr ] \ket{\Phi}
   \\ \notag
   & \doteq
       - \, \delta_{n_\alpha n_2} \delta_{\vec p_\beta \vec p_1} \delta_{\vec p_2 \vec p_a} \delta_{n_1 n_b}
       - \, \delta_{n_\alpha n_1} \delta_{\vec p_\beta \vec p_2} \delta_{\vec p_1 \vec p_a} \delta_{n_2 n_b}
   \>.
\end{align}

\end{itemize}

\section{($A=1$): Equations we need to solve}
\label{app:En}

In general the integral over the continuum \emph{particle} states
must be viewed as an integral over the continuum basis states,
plus a sum over the set of discrete states in the spectrum, which
remain  \emph{unoccupied} in the physical vacuum $\ket{\Phi}$. For
simplicity, in the main body of the paper we assumed that all
particle states are located in the continuum and this explicit sum
was left out. In this appendix we revisit this issue and list the
equivalent of Eqs.~(\ref{eq:psi_eig}, \ref{eq:chi_eig},
\ref{eq:P_energy}, \ref{eq:P_s1_p}, \ref{eq:energy},
\ref{eq:chi_h}), for the calculation of the energy and \nph{1}
correlations. We also list the equivalent of Eqs.~(\ref{eq:stil},
\ref{eq:expect_o}, \ref{eq:tils1}, \ref{eq:baro}) required for the
calculation of observables in momentum and coordinate space.

\subsection{Correlations and energies}

\begin{itemize}

\item Eigenstate of the hamiltonian $\h$ :
\begin{align}
   \ket{\Psi} \ = \ &
   \ket{\phi_h}
   + \sum_{n \neq h} \msb{n}{h} \ket{\phi_n}
   + \dmom{p} \msa{p}{h} \ket{\phi_{\vec p}}
   \>,
\end{align}

\item \emph{Hole}-function :
\begin{equation}
   \ket{\chi_h} \ = \
   \ket{\Psi} - \ket{\phi_h} - \sum_{n \neq h} \msb{n}{h}
   \ket{\phi_n}
   \>.
\end{equation}

\item $P$-set equations :
\begin{align}
   & E \ = \ \meh{\phi_h}{\phi_h}
   + \sum_{n \neq h} \meh{\phi_h}{\phi_n} \, \msb{n}{h}
   \notag \\ & \qquad \qquad
   + \dmom{p} \meh{\phi_h}{\phi_{\vec p}} \, \msa{p}{h}
   \>,
\end{align}
\begin{align}
   & E \, \msb{n}{h} \ = \ \meh{\phi_n}{\phi_h}
   + \sum_{n_1 \neq h} \meh{\phi_n}{\phi_{n_1}} \, \msb{n_1}{h}
   \nonumber \\ & \qquad \qquad
   + \dmom{p_1} \meh{\phi_n}{\phi_{\vec p_1}} \, \msa{p_1}{h}
   \>,
\end{align}
\begin{align}
   & E \, \msa{p}{h} \ = \ \meh{\phi_{\vec p}}{\phi_h}
   + \sum_{n_1 \neq h} \meh{\phi_{\vec p}}{\phi_{n_1}} \, \msb{n_1}{h}
   \nonumber \\ & \qquad \qquad
   + \dmom{p_1} \meh{\phi_{\vec p}}{\phi_{\vec p_1}} \, \msa{p_1}{h}
   \>.
\end{align}

\item $R$-set equations :
\begin{align}
   & E \ = \ \meh{\phi_h}{\phi_h}
   + \meh{\phi_h}{\chi_h}
   \nonumber \\ & \qquad \qquad \qquad
   + \sum_{n \neq h} \meh{\phi_h}{\phi_n} \, \msb{n}{h}
   \>,
\end{align}
\begin{align}
   & E \, \msb{n}{h} \ = \ \meh{\phi_n}{\phi_h}
   + \meh{\phi_n}{\chi_h}
   \nonumber \\ & \qquad \qquad \qquad
   + \sum_{n_1 \neq h} \meh{\phi_n}{\phi_{n_1}} \, \msb{n_1}{h}
   \>,
\end{align}
\begin{align}
   & ( \h - E ) \, \ket{\chi_h}
   \ = \
   \nonumber \\ & \quad
   - \,
   ( \h - E ) \,
   \Bigl [
      \ket{\phi_h}
      + \sum_{n_1 \neq h} \ket{\phi_{n_1}} \, \msb{n_1}{h}
   \Bigr ]
   \>.
\end{align}

\end{itemize}

\subsection{Observables}

\begin{itemize}

\item $P$-set complement :
\begin{align}
   &
   \mtb{n}{h} \ = \
   \msb{n}{h} \
   \Bigl [
      1
      -
      \sum_{n_ \neq h} \mtb{n_1}{h} \msb{n_1}{h}
   \\ \notag & \qquad \qquad \qquad \qquad
      -
      \dmom{p_1} \mta{p_1}{h} \msa{p_1}{h}
   \Bigr ]
   \>,
\end{align}
\begin{align}
   &
   \mta{p}{h} \ = \
   \msa{p}{h} \
   \Bigl [
      1
      -
      \sum_{n_ \neq h} \mtb{n_1}{h} \msb{n_1}{h}
   \\ \notag & \qquad \qquad \qquad \qquad
      -
      \dmom{p_1} \mta{p_1}{h} \msa{p_1}{h}
   \Bigr ]
   \>,
\end{align}
\begin{align}
   &
   \bar o \ = \
   \me{\phi_h}{\Op}{\Psi}
   \\ \notag & \qquad \quad
      + \sum_{n \neq h} \mtb{n}{h} \Bigl [
      \me{\phi_{n}}{\Op}{\Psi}
      -
      \me{\phi_h}{\Op}{\Psi}
      \msb{n}{h}
   \\ \notag & \qquad \quad
      + \dmom{p} \mta{p}{h} \Bigl [
      \me{\phi_{\vec p}}{\Op}{\Psi}
      -
      \me{\phi_h}{\Op}{\Psi}
      \msa{p}{h}
   \Bigr ]
   \>.
\end{align}

\item $R$-set complement :
\begin{align}
   &
   \mtb{n}{h} \ = \
   \msb{n}{h} \
   \Bigl [
      1
      -
      \frac{1}{E} \, \meh{\tilde \chi_h}{\Psi}
   \\ \notag & \qquad \qquad \qquad \qquad
      -
      \sum_{n_1 \neq h} \mtb{n_1}{h}
      \frac{1}{E} \, \meh{\phi_{n_1}}{\Psi}
   \Bigr ]
   \>,
\end{align}
\begin{align}
   &
   \bra{\tilde \chi_h} \ = \
   \bra{\chi_h} \
   \Bigl [
      1
      -
      \frac{1}{E} \, \meh{\tilde \chi_h}{\Psi}
   \\ \notag & \qquad \qquad \qquad \qquad
      -
      \sum_{n_1 \neq h} \mtb{n_1}{h}
      \frac{1}{E} \, \meh{\phi_{n_1}}{\Psi}
   \Bigr ]
   \>,
\end{align}
\begin{align}
   &
   \bar o \ = \
   \me{\phi_h}{\Op}{\Psi}
   + \me{\tilde \chi_h}{\Op}{\Psi}
   - \frac{1}{E} \, \meh{\tilde \chi_h}{\Psi}
     \me{\phi_h}{\Op}{\Psi}
   \notag \\ & \qquad
   + \sum_{n \neq h}
     \Bigl [ \me{\phi_n}{\Op}{\Psi}
             - \frac{1}{E} \, \meh{\phi_n}{\Psi}
                              \me{\phi_h}{\Op}{\Psi}
     \Bigr ]
   \>.
\end{align}

\end{itemize}

\section{$R$-set practical implementation}
\label{app:cce_a1}

Consider two complete sets of functions, $\ket{\phi_n}$ and
$\ket{\psi_n}$. We model the physical vacuum using elements of the
first set, while the second set is used to determine the
corresponding hole-function. In general, we expect $\ket{\phi_n}$
to offer a good description for the interior part of the wave
function, while $\ket{\psi_n}$ will provide a good description for
the tail of the wave function. In practice, we use harmonic
oscillator and Laguerre wave functions as the set $\ket{\phi_n}$
and the set $\ket{\psi_n}$, respectively. Note that a different
viable candidate for the second basis may be the set of
transformed harmonic oscillator wave functions introduced
in~\cite{ref:SNP}, functions which are obtained via a local-scale
point transformation of the spherical harmonic oscillator basis.
At this time we find the Laguerre functions simpler to handle.
Properties of the Laguerre functions are reviewed in
Appendix~\ref{app:wfcn}.

We list now the necessary equations for the ground state.
Extension of the formalism to the excited states calculation
follows without difficulty. For the ground state, we take
$\ket{\Phi} = \ket{\phi_1}$. We write the hole-function as an
expansion in the second set of functions:
\begin{equation}
   \ket{\chi_1} \ = \
   \sum_n \ c_n \ \ket{\bar \psi_n} \>,
\end{equation}
with $\ket{\bar \psi_n} = \ket{\psi_n} -
\ket{\phi_1}\dotp{\phi_1}{\psi_n}$. Since $\ket{\psi_n}$ are not
orthogonal to $\ket{\phi_1}$, the overlap substraction is
necessary to fulfill the condition $\dotp{\chi_1}{\phi_1} = 0$.
The price we have to pay is that the set of functions $\ket{\bar
\psi_n}$ are no longer orthogonal to each other. The expansion
coefficients $c_n$ are found as the solution of the linear system
of equations~\eqref{eq:chi_h} :
\begin{align}
   \sum_n \ c_n \ \Bigl [ \meh{\bar \psi_m}{\bar \psi_n} \ - \ E
   \, \dotp{\bar \psi_m}{\bar \psi_n} \Bigr ]
   \ = \ - \ \meh{\bar \psi_m}{\phi_1}
   \>,
\end{align}
where
\begin{align}
   &
   \meh{\bar \psi_m}{\bar \psi_n}
   \ = \
   \meh{\psi_m}{\psi_n}
   + E_1 \, \dotp{\psi_m}{\phi_1} \dotp{\phi_1}{\psi_n}
   \\ \notag & \qquad \qquad
   - \dotp{\psi_m}{\phi_1} \meh{\phi_1}{\psi_n}
   - \meh{\psi_m}{\phi_1} \dotp{\phi_1}{\psi_n}
   \>,
\end{align}
and
\begin{align}
   \dotp{\bar \psi_m}{\bar \psi_n}
   \ = \
   \delta_{mn} \ - \
   \dotp{\psi_m}{\phi_1} \dotp{\phi_1}{\psi_n}
   \>.
\end{align}
We calculate the ground state energy by using
Eq.~\eqref{eq:energy} :
\begin{align}
   E &
   \ = \
   \meh{\phi_1}{\phi_1}
   \ + \ \sum \ c_n \ \meh{\phi_1}{\bar \psi_n} \>,
\end{align}
where
\begin{align}
   \meh{\phi_1}{\bar \psi_n}
   \ = \
   \meh{\phi_1}{\psi_n} - E_1 \, \dotp{\phi_1}{\psi_n}
   \>.
\end{align}

For the calculation of observables we also need the modified
hole-function, $\ket{\tilde \chi_1}$. Once again we perform an
expansion in terms of the set $\ket{\psi_n}$. We have
\begin{equation}
   \ket{\tilde \chi_1} \ = \
   \sum_n \ d_n \ \ket{\bar \psi_n} \>,
\end{equation}
where the expansion coefficients are found by solving the linear
system of equations~\eqref{eq:tils1} :
\begin{align}
   \sum_n \ d_n \
   \Bigl [ \dotp{\bar \psi_n}{\bar \psi_m}
           +
           \frac{1}{E} \ \dotp{\chi_1}{\bar \psi_m} \ \meh{\bar \psi_n}{\Psi}
   \Bigr ] =
   \dotp{\chi_1}{\bar \psi_m}
   \>.
\end{align}
Subsequently, observables are calculated using
Eq.~\eqref{eq:baro}. For \emph{local} operators, it is sufficient
to calculate the one-body density
\begin{align}
   \rho(\vec r) & \ = \
   \me{\Psi}{\delta(\vec r - \vec r_1}{\Psi} / \dotp{\Psi}{\Psi}
   \\ \notag & \ = \
   \phi_1^*(\vec r) \Psi(\vec r)
   \Bigl [ 1 - \frac{1}{E} \, \meh{\tilde \chi_1}{\Psi} \Bigr ]
   \ + \
   \tilde \chi_1^*(\vec r) \Psi(\vec r)
   \>.
\end{align}
Then, the expectation value of an arbitrary one-body local
operator $\mathcal{O}$ can be calculated as
\begin{equation}
   \me{\Psi}{\mathcal{O}(\vec r)}{\Psi} \ = \
   \dpos{r_1} \ \mathcal{O}(\vec r_1) \
   \me{\Psi}{\delta(\vec r - \vec r_1}{\Psi}
   \>.
\end{equation}

\section{Single-particle wave functions}
\label{app:wfcn}

We briefly review here the properties of the single-particle wave
functions employed in the numerical part of this paper. We follow
the conventions of Ref.~\cite{ref:arfken}.
\begin{itemize}
   \item the harmonic oscillator functions have the form
\begin{align}
   &
   \mathcal{HO}_{n\ell}(x) \ \sim \
   x^{\ell+1} e^{- x^2 / 2} \
   L_n^{\ell+1/2}(x^2)
   \>,
\end{align}
with $x = r/b$, and obey the orthogonality condition
\begin{align}
   \int_0^\infty \mathrm{d}r \ \mathcal{HO}_{n\ell}(r) \mathcal{HO}_{m\ell}(r)
   =
   \frac{(2 n + 2 \ell + 1)!! \sqrt{\pi}}{n! \ b^3 \ 2^{n+\ell+1}}
   \ \delta_{nm}
   \>.
\end{align}
Here $\{ L_n^k(x) \, ; \ n = 0,1,\ldots \}$ are the associated
Laguerre polynomials.

For the calculation of the kinetic energy matrix element we will
need the identity
\begin{align}
   &
      \left [
       \frac{\mathrm{d}^2}{\mathrm{d}r^2}
       - \frac{\ell(\ell+1)}{r^2}
   \right ] \ \mathcal{HO}_{n\ell}(r)
   \\ \notag
   & \ = \
   \frac{1}{b^2} \
   \Bigl [ 2 ( 2n + \ell + \frac{3}{2} )
           - x^2
   \Bigr ] \
   \mathcal{HO}_{n\ell}(x)
   \>,
\end{align}
and the nonzero elements of the $\langle n \ell | x | m
\ell\rangle$ matrix
\begin{align}
   \me{n \ell}{x^2}{n \ell}
   \ = \
   2n + \ell + \frac{3}{2} \>,
   \\
   \me{n-1 \ \ell}{x^2}{n \ell}
   \ = \
   - \
   \sqrt{ n (n+\ell+\frac{1}{2}) }
   \>.
\end{align}

   \item the Laguerre functions have the form
\begin{equation}
   \mathcal{L}_n^{2\ell+2}(x) \ \sim \
   x^{\ell+1} e^{-x/2} \ L_n^{2\ell+2}(x)
   \>,
\end{equation}
where $x = \gamma r$, and obey the orthogonality condition
\begin{align}
   \int_0^\infty \mathrm{d}x \
       \mathcal{L}_n^{2\ell+2}(x) \ \mathcal{L}_m^{2\ell+2}(x)
   \ = \ &
   \frac{(n+2\ell+2)!}{n!} \ \delta_{nm}
   \>.
   \notag \\
\end{align}
For the calculation of the kinetic energy matrix element we need
the identity
\begin{align}
   &
   \left [
       \frac{\mathrm{d}^2}{\mathrm{d}x^2}
       - \frac{\ell(\ell+1)}{x^2}
   \right ] \ \mathcal{L}_n^{2\ell+2}(x)
   \\ \notag &
   \ = \
   \left (
      \frac{1}{4}
      - \frac{n+\ell+1}{x} - \frac{n}{x^2}
   \right ) \ \mathcal{L}_n^{2\ell+2}(x)
   \\ \qquad & \qquad
   \ + \
   \frac{n+2\ell+2}{x^2} \ \mathcal{L}_{n-1}^{2\ell+2}(x)
   \>.
\end{align}
The above equation involves the \emph{unnormalized} Laguerre
functions. If the normalized Laguerre functions are desired, then
the coefficient of $\mathcal{L}_{n-1}^{2\ell+2}$ must be modified
accordingly.

The following identities pertain to the matrix element calculation
of $\langle n \ell | x^{-p} | m \ell \rangle$, with $p=1,2$. We
have
\begin{align}
   &
   \int_0^\infty \mathrm{d}x \ x^{2\ell+1} \ e^{-x} \
   L_n^{2\ell+2}(x) L_m^{2\ell+2}(x)
   \\ \notag &
   \ = \
   \sum_{c=0}^{\mathrm{min}\{n,m\} } \
   \frac{ (2\ell+1 + c)\, !}{c\,!} \
   \>,
\end{align}
and
\begin{align}
   &
   \int_0^\infty \mathrm{d}x \ x^{2\ell} \ e^{-x} \
   L_n^{2\ell+2}(x) L_m^{2\ell+2}(x)
   \\ \notag &
   \ = \
   \sum_{c=0}^{\mathrm{min}\{n,m\} }
      (n + 1 - c) (m + 1 - c) \,
   \frac{ (2\ell + c)\, !}{c\,!}
   \>.
\end{align}

\end{itemize}

\vfill
%
%

\end{document}